\documentclass[10pt]{IEEEtran}

\usepackage{amsmath}
\usepackage{amsthm}
\usepackage{amsfonts}
\usepackage{graphicx}
\usepackage{cite}

\newtheorem{theorem}{Theorem}
\newtheorem{lemma}{Lemma}
\newtheorem{remark}{Remark}
\allowdisplaybreaks

\newcommand{\ber}	{\begin{eqnarray*}}
\newcommand{\eer}	{\end{eqnarray*}}

\newcommand{\eps}	{\epsilon}

\newcommand{\cX}	{{\cal X}}
\newcommand{\cY}	{{\cal Y}}

\newcommand{\cP}	{{\cal P}}

\newcommand{\tilX}{\tilde{X}}
\newcommand{\tilY}{\tilde{Y}}

\usepackage{color}

\newcounter{ho}
\newcounter{lec}
\newcounter{hw}
\newcounter{hs}

\DeclareMathOperator*{\argmax}{arg\,max}

\title{Every Bit Counts: Second-Order Analysis of Cooperation in the Multiple-Access Channel}

\author{Oliver Kosut, Michelle Effros, Michael Langberg
\thanks{O. Kosut is with the School of Electrical, Computer and Energy Engineering at Arizona State University. Email: {\tt okosut@asu.edu}}
\thanks{M. Effros is with the Department of Electrical Engineering at the California Institute of Technology.
Email: \texttt{effros@caltech.edu}}
\thanks{M. Langberg is with the Department of Electrical Engineering at the University at Buffalo (State University of New York).  
Email: \texttt{mikel@buffalo.edu}}
\thanks{This work is supported in part by NSF grants CCF-1817241, CCF-1908725, and CCF-1909451. 
}
}

\begin{document}

\maketitle

\begin{abstract}
The work at hand presents a finite-blocklength analysis of the multiple access channel (MAC) sum-rate under the cooperation facilitator (CF) model. 
The CF model, in which independent encoders coordinate through an intermediary node, is known to show significant rate benefits, even when the rate of cooperation is limited.
We continue this line of study for cooperation rates which are sub-linear in the blocklength $n$. 
Roughly speaking, our results show that if the facilitator transmits $\log{K}$ bits, there is a sum-rate benefit of order $\sqrt{\log{K}/n}$. This result extends across a wide range of $K$: even a single bit of cooperation is shown to provide a sum-rate benefit of order $1/\sqrt{n}$.
\end{abstract}

\section{Introduction}

The multiple access channel (MAC) model 
lies at an interesting conceptual intersection 
between the notions of cooperation and interference 
in wireless communications.
When viewed from the perspective 
of any single transmitter, 
codewords transmitted by other transmitters 
can only inhibit the first transmitter's 
{\em individual} communication rate;  
thus each transmitter sees the others 
as a source of interference.  
When viewed from the perspective of the receiver, 
however, 
maximizing the {\em total} rate 
delivered to the receiver 
often requires 
all transmitters to communicate simultaneously; 
from the receiver's perspective, then, 
the transmitters must cooperate 
through their simultaneous transmissions
to maximize the sum-rate 
delivered to the receiver. 

Simultaneous transmission 
is, perhaps, the weakest form of cooperation imaginable 
in a wireless communication model.  
Nonetheless, the fact that 
even simultaneous transmission 
of independent codewords 
from interfering transmitters 
can increase the sum-rate 
deliverable to the MAC receiver 
begs the question of how much more could be achieved 
through more significant forms 
of MAC transmitter cooperation.  

The information theory literature 
devotes considerable effort 
to studying the impact of encoder cooperation 
in the MAC.  
A variety of cooperation models are considered.  
Examples include 
the ``conferencing'' cooperation 
model~\cite{willems1983discrete}, 
in which encoders share information directly 
in order to coordinate their channel inputs, 
the ``cribbing'' cooperation model~\cite{willems1985discrete}, in which transmitters cooperate by sharing their codeword information (at times causally), 
and the ``cooperation facilitator'' (CF) 
cooperation model~\cite{noorzad2014power}
in which users coordinate their channel inputs 
with the help of an intermediary called the CF.    
The CF distinguishes the amount of information 
that must be understood to facilitate cooperation 
(i.e., the rate $R_{\rm IN}$ {\em to} the CF) 
from the amount of information 
employed in the coordination  
(i.e., the rate $R_{\rm OUT}$ {\em from} the CF).  
Key results using the CF model show 
that for many MACs, 
no matter what the (non-zero) fixed rate $C_{\rm IN}$, 
the curve 
describing the maximal sum-rate as a function 
of $R_{\rm OUT}$ has infinite slope 
at $R_{\rm OUT}=0$~\cite{NEL:18}. 
That is, {\em very little coordination through a CF
can change the MAC capacity considerably.}
This phenomenon holds for both average and maximum error sum-rates;  
it is most extreme in the latter case, where 
even a finite number of bits 
(independent of the blocklength) 
--- that is, $R_{\rm OUT}=0$ ---
can suffice to change the MAC capacity 
region~\cite{langberg2016capacity,noorzad2018can,NLE:19}.  

We study the CF model for 2-user MACs under the average error criterion.
In this setting, the maximal sum-rate is a {\em continuous} function of $R_{\rm OUT}$ at $R_{\rm OUT}=0$~\cite{noorzad2018can,NLE:19}, implying a {\em first-order} upper-bound on the benefit of  cooperation for rates that are sub-linear.
However, sub-linear CF cooperation may still increase sum-rate, albeit through second-order terms.  
In this work, we seek to understand the impact of the CF 
over a wide range of cooperation rates. Specifically, we consider a CF that, after viewing both messages, can transmit one of $K$ signals to both transmitters. We prove achievable bounds that express the benefit of this cooperation as a function of $K$. These bounds extend all the way from constant $K$ to exponential $K$. Interestingly, we find that even for $K=2$ (i.e., one bit of cooperation), there is a benefit in the second-order (i.e., dispersion) term, corresponding to an improvement of $O(\sqrt{n})$ message bits. We prove two main achievable bounds, each of which is optimal for a different range of $K$ values. The proof of the first bound is based on refined asymptotic analysis similar to typical second-order bounds. The proof of the second bound is based on the method of types. For a wide range of $K$ values, we find that the benefit is $O(\sqrt{n\log K})$ message bits.

\section{Problem Setup}

An $(M_1,M_2,K)$ facilitated multiple access code 
for multiple access channel (MAC) 
\[
(\cX_1\times\cX_2,p_{Y|X_1,X_2}(y|x_1,x_2),\cY)
\]
is defined by 
a facilitator code
\begin{eqnarray*}
& e: & [M_1]\times[M_2] \rightarrow [K]
\end{eqnarray*}
a pair of encoders 
\begin{eqnarray*}
& f_1: & [M_1]\times[K] \rightarrow \cX_1 \\
& f_2: & [M_2]\times[K] \rightarrow \cX_2 
\end{eqnarray*}
and a decoder 
\[
g:\cY\rightarrow[M_1]\times[M_2].
\]
The encoder's output is sometimes described using the abbreviated notation 
\begin{eqnarray*}
X_1(m_1,m_2) & = & f_1(m_1,e(m_1,m_2)) \\
X_2(m_1,m_2) & = & f_2(m_2,e(m_1,m_2)).
\end{eqnarray*}
The average error probability for the given code is 
\begin{align*}
P_e &= \frac1{M_1M_2}\sum_{m_1=1}^{M_1}\sum_{m_2=1}^{M_2}
\Pr\big(g(Y)\neq (m_1,m_2) \big|
\\&\qquad (X_1,X_2)=(X_1(m_1,m_2),X_2(m_1,m_2))\big).
\end{align*}
We also consider codes for the $n$-length product channel, where $\cX_1,\cX_2,\cY$ are replaced by $\cX_1^n,\cX_2^n,\cY^n$ respectively, and where
\[
p_{Y^n|X_1^n,X_2^n}(y^n|x_1^n,x_2^n)=\prod_{i=1}^n p_{Y|X_1,X_2}(y_i|x_{1i},x_{2i}).
\]
An $(M_1,M_2,K)$ code for the $n$-length channel achieving average probability of error at most $\eps$ is called an $(n,M_1,M_2,K,\eps)$ code. We assume that all alphabets are finite.

The following notation will be useful. Given a MAC 
\[(\cX_1\times\cX_2, p_{Y|X_1,X_2}(y|x_1,x_2), \cY),\]
the sum-capacity without cooperation is given by
\begin{equation}\label{sum_capacity}
C_{\text{sum}}=\max_{p_{X_1}p_{X_2}}I(X_1,X_2;Y).
\end{equation}
Let $\cP^\star$ be the set of product distributions $p_{X_1}p_{X_2}$ achieving the maximum in \eqref{sum_capacity}.
For any $p_{X_1}p_{X_2}\in\cP^\star$, let $p_Y$ be the resulting marginal on the channel output, giving
\[
p_{Y}(y) = \sum_{(x_1,x_2)\in\cX_1\times\cX_2}p_{X_1}(x_1)p_{X_2}(x_2)p_{Y|X_1,X_2}(y|x_1,x_2)
\]
for all $y\in\cY$.  
We use $i(x_1,x_2;y)$, $i(x_1;y|x_2)$ and $i(x_2;y|x_1)$ 
to represent the joint and conditional information densities
\begin{eqnarray*}
i(x_1,x_2;y) & = & \log \left(\frac{p_{Y|X_1,X_2}(y|x_1,x_2)}{p_Y(y)}\right) \\
i(x_1;y|x_2) & = & \log \left(\frac{p_{Y|X_1,X_2}(y|x_1,x_2)}{p_{Y|X_2}(y|x_2)}\right) \\
i(x_2;y|x_1) & = & \log \left(\frac{p_{Y|X_1,X_2}(y|x_1,x_2)}{p_{Y|X_1}(y|x_1)}\right), 
\end{eqnarray*}
where $p_{Y|X_1}$ and $p_{Y|X_2}$ are conditional marginals on $Y$ under joint distribution 
\[
p_{X_1,X_2,Y}=p_{X_1}p_{X_2}p_{Y|X_1,X_2}.
\]
We denote the 3-vector of all three quantities as
\[
\mathbf{i}(x_1,x_2;y)=\left[\begin{array}{c}i(x_1,x_2;y)\\ i(x_1;y|x_2)\\ i(x_2;y|x_1)\end{array}\right].
\]
It will be convenient to define
\begin{align*}
i(x_1,x_2)&=E[i(x_1,x_2;Y)|(X_1,X_2)=(x_1,x_2)]
\\&=D(p_{Y|X_1=x_1,X_2=x_2}\|p_Y).
\end{align*}
Let
\begin{align}
V_1&=\text{Var}(i(X_1,X_2)),\label{sigma_def}
\\
V_2 &= E[\text{Var}(i(X_1,X_2;Y)|X_1,X_2)].\label{V_def}
\end{align}
Roughly speaking, $V_1$ represents the information-variance of the codewords, whereas $V_2$ represents the information-variance of the channel noise. Given two distributions $p_X,q_X$, let the divergence-variance be
\[
V(p_X\|q_X)=\text{Var}_{p_X}\left(\log \frac{p_X(X)}{q_X(X)}\right).
\]
Note that
\[
V_2=\sum_{x_1,x_2} p_{X_1}(x_1)p_{X_2}(x_2)V(p_{Y|X_1=x_1,X_2=x_2}\|p_Y).
\]

\section{Main Results}

Define the fundamental sum-rate limit for the facilitated-MAC as
\begin{multline*}
R_{\text{sum}}(n,\eps,K)\\ =\sup\bigg\{\frac{\log(M_1M_2)}{n}:
 \exists (n,M_1,M_2,K,\eps)\text{ code}\bigg\}.
\end{multline*}

In the literature on second-order rates, there are typically two types of results: (i) finite blocklength results, with no asymptotic terms, that are typically written in terms of abstract alphabets, and (ii) asymptotic results that derive from these finite blocklength results, which are typically easier to understand. The following is an achievable result which has some flavor of both: the channel noise is dealt with via an asymptotic analysis, but the dependence on the randomness in the codewords is written as in a finite blocklength result. We provide this ``intermediate'' result because, depending on the CF parameter $K$, the relevant aspect of the codeword distribution may be in the central limit, moderate deviations, or large deviations regime. Thus, in this form one may plug in any concentration bound to derive an achievable bound. Subsequently,  Theorem~\ref{thm:small_deviations} gives specific achievable results based on two different concentration bounds. We also prove another achievable bound, Theorem~\ref{thm:large_deviations}, which does not rely on Theorem~\ref{thm:initial_bound}, but instead uses an approach based on the method of types that applies at larger values of $K$.

\begin{theorem}\label{thm:initial_bound}
Assume $\log K=o(n)$. For any distribution $p_{X_1},p_{X_2}$, let $X_j^n(k)$ be an i.i.d.\ sequence from $p_{X_j}$ for each $k\in[K]$, with all sequences mutually independent. There exists an $(n,M_1,M_2,K,\eps)$ code if
\begin{align}
\eps &\ge \Pr\Bigg(\max_{k\in[K]}\sum_{i=1}^n i(X_{1i}(k),X_{2i}(k))+\sqrt{nV_2}\,Z_0\notag
\\&\qquad<\log(M_1M_2K)+\frac{1}{2}\log n\Bigg)\notag
\\&\qquad+O\left(\sqrt{\frac{\log n}{n}}\right)+O\left(\sqrt{\frac{\log K}{n}}\right)\label{eps_condition}
\\
\log M_1&\le nI(X_1;Y|X_2)-c\sqrt{n\log K+n\log n}\label{M1_condition}
\\\log M_2&\le nI(X_2;Y|X_1)-c\sqrt{n\log K+n\log n}\label{M2_condition}
\end{align}
where $Z_0$ is a standard Gaussian, and where $c$ is a constant.
\end{theorem}

\begin{figure}
\begin{center}
\includegraphics[width=.8\columnwidth]{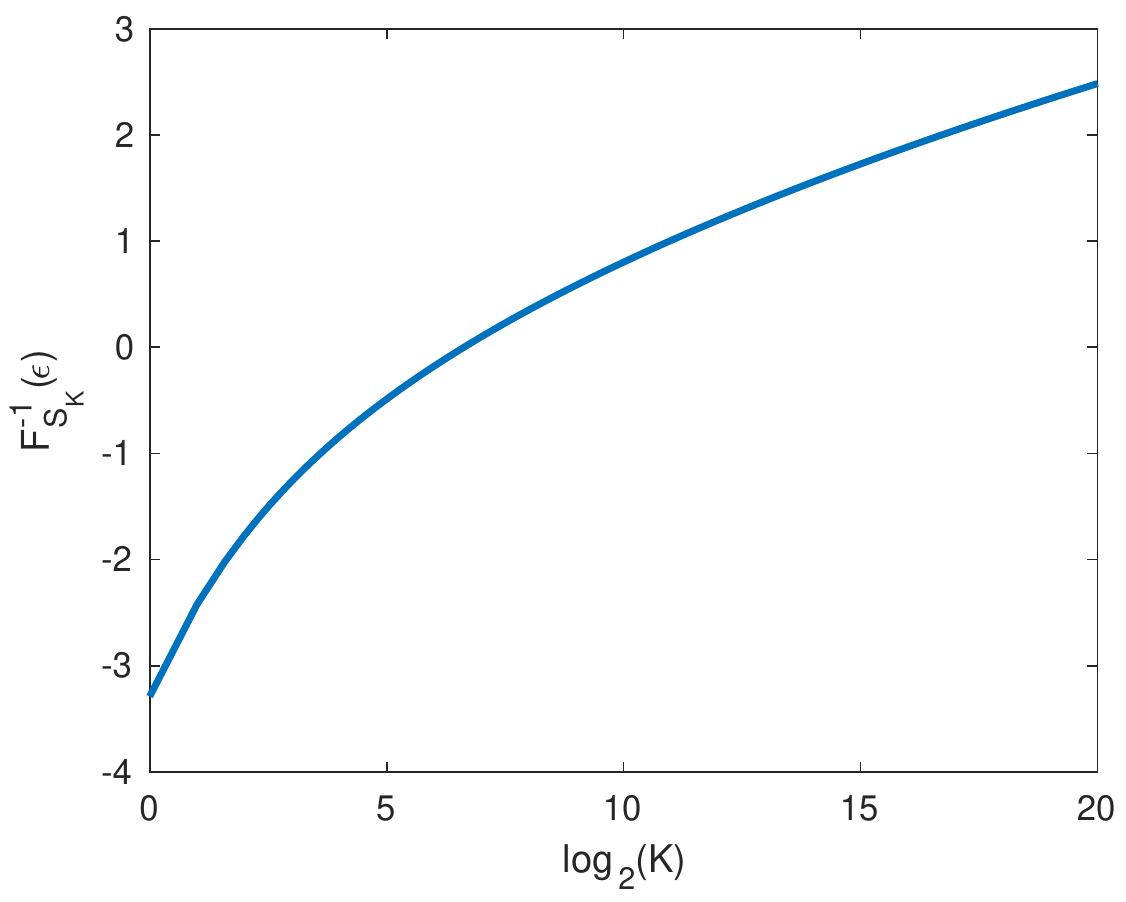}
\end{center}
\caption{The inverse CDF $F_{S_K}^{-1}(\eps)$ for $\eps=0.01$, for $V_1=V_2=1$ across a range of $K$. Note that the horizontal axis is $\log_2 K$, i.e., the number of bits transmitted from the CF.}
\label{fig:CDF_inverse}
\end{figure}

For fixed $K$, let $Z_0,\ldots,Z_{K}$ be drawn i.i.d. from $\mathcal{N}(0,1)$. Let
\[
S_K = \sqrt{V_2}\,Z_{0}+ \sqrt{V_1} \max_{k\in[1:K]}Z_k,
\]
and define the CDF of $S_K$ as
\[
F_{S_K}(s)=\Pr(S_K\le s).
\]
Also let $F_{S_K}^{-1}$ be the inverse of the CDF;  that is,
\[
F_{S_K}^{-1}(p)=\sup\{s:F_{S_K}(s)\le p\}\text{ for }p\in[0,1].
\]
In what follows we use Theorem~\ref{thm:initial_bound} and the function $F_{S_K}^{-1}$ to explicitly bound from below the benefit in sum-rate when cooperating with varying measures of $K$. A numerical computation of $F_{S_K}^{-1}(\eps)$ as a function of $K$ is shown in Fig.~\ref{fig:CDF_inverse}. The following is a technical estimate of $F_{S_K}^{-1}$.

\begin{lemma}
\label{lem:max_gausians}
For $K$ and $\eps$ that satisfy $K>e^3\sqrt{2\pi}\ln(4/\eps)$, $F_{S_K}^{-1}(\eps)$ is at least
\begin{multline*}
\sqrt{V_1(2\ln K-2\ln\ln({4}/{\eps})-\ln\ln K-\ln(4\pi))}\\ -\sqrt{2V_2\ln(2/\eps)}.
\end{multline*}
Moreover, for all $K$ and $\eps$,
\[
F^{-1}_{S_K}(1-\eps) \leq \sqrt{2V_1\ln K}+ \sqrt{2V_1\ln(4/\eps)}+\sqrt{2V_2\ln(2/\eps)}.
\]
\end{lemma}

\begin{IEEEproof}
Let $Z(K)=\max_{k\in[K]}{Z_k}$.
From \cite{hartigan2014bounding},
it holds that
$\Pr(Z(K) \leq \sqrt{\kappa-\ln{\kappa}}) \leq \eps/2$ 
for $\kappa = 2\ln(K/\sqrt{2\pi})-2\ln\ln(4/\eps) \geq 6$.
Moreover, $\Pr(\sqrt{V}Z_0 \leq -\sqrt{2V\ln(2/\eps)}) \leq \eps/2$.
Combining these bounds gives the desired lower bound.

For the upper bound,  \cite{Borell,TIS} 
imply that for any $K$, $\Pr(\sqrt{V_1} Z(K)\geq (\sqrt{2V_1\ln K}+ \sqrt{2V_1\ln(4/\eps)}))\leq \eps/2$.
Moreover, $\Pr(\sqrt{V_2}Z_0 \geq \sqrt{2V_2\ln(2/\eps)}) \leq \eps/2$.
Thus,
$F^{-1}_{S_k}(1-\eps) \leq \sqrt{2V_1\ln K}+ \sqrt{2V_1\ln(4/\eps)}+\sqrt{2V_2\ln(2/\eps)}$.
\end{IEEEproof}

\begin{theorem}\label{thm:small_deviations}
For any $p_{X_1}p_{X_2}\in\cP^\star$ and the associated constants $V_1$ and $V_2$, if $\log K=o(n^{1/3})$, then
\[
R_{\text{sum}}(n,\eps,K)\ge C_{\text{sum}}+\frac{1}{\sqrt{n}}\,F_{S_K}^{-1}(\eps)-\theta_n
\]
where
\begin{align}
\theta_n&=O\left(\frac{\log n}{n}\right),&&\text{ if }K\le\log n\label{theta1}\\
\theta_n&=O\left(\frac{K}{n}\right),&&\text{ if }\log n\le K\le \log^{3/2}n\label{theta2}\\
\theta_n&=O\left(\frac{\log^{3/2}n}{n}\right),&&\text{ if }\log^{3/2}n\le K\le n\label{theta3}\\
\theta_n&=O\left(\frac{\log^{3/2}K}{n}\right),&&\text{ if }K\ge n.\label{theta4}
\end{align}
\end{theorem}

For larger $K$, our achievability bound employs the function
\[
\Delta(a)=\max_{p_{X_1,X_2}:I(X_1;X_2)\le a} I(X_1,X_2;Y)-C_{\text{sum}}.
\]
Note that $\Delta(0)=0$. Lemma~\ref{large_but_small} captures the behavior of $\Delta(a)$ for small $a$. (See Appendix~\ref{appendix:large_but_small} for the proof.)

\begin{lemma}\label{large_but_small}
In the limit as $a\to 0$,
\[
\Delta(a)=\sqrt{a\,2V_1^\star\ln 2}+o(\sqrt{a})
\]
where 
\begin{equation}\label{Vstar_def}
V_1^\star=\max_{p_{X_1}p_{X_2}\in\cP^\star} \text{{\em Var}}(i(X_1,X_2)).
\end{equation}
\end{lemma}

\begin{theorem}\label{thm:large_deviations}
For any $K$ such that $\log K=\omega(\log n)$,
\begin{multline*}
R_{\text{sum}}(n,\eps,K)\\
\ge C_{\text{sum}}+\Delta\left(\frac{\log K}{n}-O\left(\frac{\log n}{n}\right)\right)
-O\left(\frac{1}{\sqrt{n}}\right).
\end{multline*}
\end{theorem}

\begin{remark}
While Theorems~\ref{thm:small_deviations} and~\ref{thm:large_deviations} appear quite different, Lemmas~\ref{lem:max_gausians} and~\ref{large_but_small} imply  that for mid-range $K$ values, they give  similar results. In particular, if
\[
\log n \ll \log K \ll n^{1/3}
\]
then applying Theorem~\ref{thm:small_deviations}, and choosing the distribution $p_{X_1}p_{X_2}\in\cP^\star$ that achieves the maximum in \eqref{Vstar_def} gives
\begin{align*}
R_{\text{sum}}(n,\eps,K)-C_{\text{sum}}&\ge\frac{1}{\sqrt{n}} F_{S_K}^{-1}(\eps)-\theta_n
\\&\approx \sqrt{\frac{2V_1^\star\ln K}{n}}.
\end{align*}
For the same range of $K$, Theorem~\ref{thm:large_deviations} gives
\begin{align*}
R_{\text{sum}}(n,\eps,K)-C_{\text{sum}}
&\ge \Delta\left(\frac{\log K}{n}-O\left(\frac{\log n}{n}\right)\right)
\\&\qquad-O\left(\frac{1}{\sqrt{n}}\right)
\\&\approx \sqrt{\frac{V_1^\star\log K}{n}\,2\ln 2}
\\&= \sqrt{\frac{2V_1^\star\ln K}{n}}.
\end{align*}
\end{remark}

\subsection{Comparison to prior work}

In \cite{NEL:18}, an analog to Theorem~\ref{thm:large_deviations} is proven for the asymptotic blocklength regime. 
Namely, in our notation, \cite{NEL:18} proves 
that for any $\eps>0$ and $\delta>0$, if we set  $K=2^{\Omega(n)}$ then there exist $n$ such that,
\[
R_{\text{sum}}(n,\eps,K)-C_{\text{sum}}> \Delta\left(\frac{\log{K}}{n}\right)-\delta.
\]
Similarly, in \cite{NEL:18,NLE:19}, an analog to Lemma~\ref{large_but_small} is shown for asymptotic blocklength. Specifically,  it is shown that the existence of distributions $p_{X_1}p_{X_2} \in \cP^\star$ and $p_{\tilde{X}_1\tilde{X}_2}$ over $\cX_1 \times \cX_2$ such that 
(a) the support of $p_{\tilde{X}_1\tilde{X}_2}$ is included in that of $p_{X_1}p_{X_2}$, and (b)
\[I(\tilde{X}_1,\tilde{X}_2,\tilde{Y})+D(p_{\tilde{X}_1\tilde{X}_2}\|p_{X_1}p_{X_2}) > I(X_1,X_2;Y)
\]
for 
\vspace{-5mm}
\begin{align*}
&p_{X_1,X_2,\tilde{X}_1,\tilde{X}_2,Y,\tilde{Y}}(x_1,x_2,\tilde{x}_1,\tilde{x}_2,y,\tilde{y})
\\&=p_{X_1}(x_1)p_{X_2}(x_2)p_{\tilde{X}_1,\tilde{X}_2}(\tilde{x}_1,\tilde{x}_2)
\\&\qquad \cdot p_{Y|X_1,X_2}(y|x_1,x_2)
p_{Y|X_1,X_2}(\tilde{y}|\tilde{x}_1,\tilde{x}_2),
\end{align*}
imply that there exists a constant $\sigma_0$ such that 
\[
\lim_{a \rightarrow 0}\Delta(a) \geq \sigma_0\sqrt{a}.
\]
Although Theorem~\ref{thm:large_deviations} and Lemma~\ref{large_but_small} (and their proof techniques) are similar in nature to those of \cite{NEL:18,NLE:19}, the analysis presented here is refined in that it captures higher order behavior in blocklength $n$ and further optimized to address the challenges in studying values of $K$ that are sub-exponential in the blocklength $n$.

We may also compare our results against prior achievable bounds without cooperation. Note that the standard MAC, with no cooperation, corresponds to $K=1$. In fact, in this case Theorem~\ref{thm:small_deviations} gives the same second-order term as the best-known achievable bound for the MAC sum-rate \cite{huang2012finite,
jazi2012simpler,
tan2014dispersions,
scarlett2015second,
yavas2020random}. This can be seen by noting that $S_1\sim\mathcal{N}(0,V_1+V_2)$, and so $F_{S_1}^{-1}(\eps)=\sqrt{V_1+V_2}\Phi^{-1}(\eps)$. Thus Theorem~\ref{thm:small_deviations} gives
\[
R_{\text{sum}}(n,\eps,1)\ge C_{\text{sum}}+\sqrt{\frac{V_1+V_2}{n}}\Phi^{-1}(\eps)-O\left(\frac{\log n}{n}\right).
\]
Moreover,
\[
V_1+V_2=\text{Var}(i(X_1,X_2;Y))
\]
which, for the optimal input distribution, is precisely the best-known achievable dispersion. The proof of Theorem~\ref{thm:small_deviations} uses i.i.d. codebooks, which, as shown in \cite{scarlett2015second}, can be outperformed in terms of second-order rate by constant combination codebooks. However, as pointed out in \cite[Sec.~III-B]{yavas2020random}, the two approaches give the same bounds on the sum-rate itself.

Another interesting conclusion comes from comparing the no cooperation case ($K=1$) with a \emph{single bit} of cooperation ($K=2$). As long as $V_1^\star>0$, it is easy to see that  $F_{S_2}^{-1}(\eps)>F_{S_1}^{-1}(\eps)$ for any $\eps\in(0,1)$ (Fig.~\ref{fig:CDF_inverse} shows an example). Thus, the second-order coefficient in Theorem~\ref{thm:small_deviations} for $K=2$ is strictly improved compared to $K=1$. Therefore, even a single bit of cooperation allows for $O(\sqrt{n})$ additional message bits.

\section{Proof of Theorem~\ref{thm:initial_bound}}

We use random code design, beginning with  independent design 
of the codewords for both transmitters.  
Precisely, we draw 
\begin{eqnarray*}
f_1(1,1),f_1(1,2),\ldots,f_1(M_1,K) & \sim & \mbox{i.i.d. }p_{X_1} \\
f_2(1,1),f_2(1,2),\ldots,f_2(M_2,K) & \sim & \mbox{i.i.d. }p_{X_2}.  
\end{eqnarray*}
The facilitator code $e(m_1,m_2)$ is then designed in an attempt to maximize the likelihood $p_{Y|X_1,X_2}$ 
under a received channel output $Y$. 
We begin by defining the threshold decoder $g(y)$ employed in our analysis. Maximum likelihood decoding is expected to give the best performance, but instead we here employ a threshold decoder for simplicity.  
For notational efficiency, let 
\begin{align*}
&(X_1,X_2)(m_1,m_2)=(X_1(m_1,m_2),X_2(m_1,m_2))
\\&=(f_1(m_1,e(m_1,m_2)),f_2(m_2,e(m_1,m_2))),
\end{align*}
where $e(m_1,m_2)$ is the (fixed) facilitator function to be defined below. Given a constant vector $\mathbf{c}^\star=[c_{12}^\star,c_1^\star,c_2^\star]^T$, we define the decoder $g(y)$ to choose the unique message pair $(m_1,m_2)$ such that
\[
\mathbf{i}((X_1,X_2)(m_1,m_2);y)\ge \mathbf{c}^\star,
\]
where the vector inequality means that all three inequalities must hold simultaneously. Conversely we use the notation $\not\ge$ between vectors to mean that any one of the three inequalities fails. If the number of message pairs that meet this constraint is not one, we declare an error. In an attempt to ensure that $i((X_1,X_2)(m_1,m_2);Y)$ is, in some sense, large 
for random channel outputs $Y$ 
that may result from that codeword pair's transmissions, 
for each $(m_1,m_2)\in[M_1]\times[M_2]$, we define 
\[
e(m_1,m_2)=\argmax_{k\in[K]}
s(f_1(m_1,k),f_2(m_2,k)),
\]
where $s(x_1,x_2)$ is a score function to be chosen below.

Under this code design, the expected error probability satisfies
\[
\begin{array}{r@{}l}
\multicolumn{2}{c}{E[P_e]
 =  E\left[\Pr\left(\left.g(Y)\neq (1,1)\right|(X_1,X_2)=(X_1,X_2)(1,1)\right)\right]}\\
\le \Pr\Bigg( 
&\mathbf{i}((X_1,X_2)(m_1,m_2);Y)\not\ge \mathbf{c}^\star
\\&\text{ or }
\mathbf{i}(X_1(\hat{m}_1,\hat{k}),X_2(\hat{m}_2,\hat{k});Y)\ge \mathbf{c}^\star 
\\&\text{ for any }(\hat{m}_1,\hat{m}_2)\neq(1,1), k\in[K]
\\& \Bigg|(X_1,X_2)=(X_1,X_2)(1,1)\Bigg).
\end{array}
\]
To further upper bound the error probability, we define the following random variables. Let $p_{\tilde{X}_1,\tilde{X}_2}$ be the joint distribution of $(X_1,X_2)(1,1)$ that results from our choice of CF. This distribution would be the same for any message pair. Let variables $X_1$, $X_2$, $\tilde{X}_1$, $\tilde{X}_2$, $Y$, $\tilde{Y}$ have joint distribution
\begin{align*}
&p_{X_1,X_2,\tilde{X}_1,\tilde{X}_2,Y,\tilde{Y}}(x_1,x_2,\tilde{x}_1,\tilde{x}_2,y,\tilde{y})
\\&=p_{X_1}(x_1)p_{X_2}(x_2)p_{\tilde{X}_1,\tilde{X}_2}(\tilde{x}_1,\tilde{x}_2)
\\&\qquad \cdot p_{Y|X_1,X_2}(y|x_1,x_2)
p_{Y|X_1,X_2}(\tilde{y}|\tilde{x}_1,\tilde{x}_2).
\end{align*}
Under transmission of message pair $(1,1)$, $(X_1,X_2,Y)$ capture the relationship between channel inputs and output in a standard MAC, whereas $(\tilde{X}_1,\tilde{X}_2,\tilde{Y})$ capture the corresponding relationship with CF. Moreover, $(\tilde{X}_1,X_2,\tilde{Y})$, $(X_1,\tilde{X}_2,\tilde{Y})$, and $(X_1,X_2,\tilde{Y})$ capture the relationship between the channel output and one or more untransmitted codewords from our random code. Assume without loss of generality that $e(1,1)=1$; i.e., $(X_1,X_2)(1,1)=(f_1(1,1),f_2(1,1))$. We now analyze the error probability by considering the following cases.

\begin{center}
\begin{tabular}{ccccc}
$\hat{m}_1$ & $\hat{m}_2$ & $\hat{k}$ & Number of values &  Distribution\\
\hline
$\ne 1$ & 1 & 1 & $M_1-1$ & $p_{X_1} p_{\tilX_2,\tilY}$\\
$\ne 1$ & 1 & $\ne 1$ & $(M_1-1)(K-1)$ & $p_{X_1} p_{X_2} p_{\tilY}$\\
1 & $\ne 1$ & 1 & $M_2-1$ & $p_{\tilX_1,\tilY} p_{X_2}$\\
$1$ & $\ne 1$ & $\ne 1$ & $(M_2-1)(K-1)$ & $p_{X_1} p_{X_2} p_{\tilY}$\\
$\ne 1$ & $\ne 1$ & any & $(M_1-1)(M_2-1)K$ & $p_{X_1} p_{X_2} p_{\tilY}$
\end{tabular}
\end{center}
Note that we have excluded cases where $\hat{m}_1=\hat{m}_2=1$, since those are not errors (even if $\hat{k}\ne 1$). Moreover, the number of cases wherein $X_1(\hat{m}_1,\hat{k}),X_2(\hat{m}_2,\hat{k}),Y$ has joint distribution  $p_{X_1} p_{X_2} p_{\tilY}$ is less than $M_1M_2K$. We can  upper bound the expected error probability as
\begin{align*}
E[P_e]
&\le \Pr\left(\mathbf{i}(\tilde{X}_1,\tilde{X}_2;\tilde{Y})\not\ge\mathbf{c}^\star\right)
\\&\quad+M_1M_2K\Pr(\mathbf{i}(X_1,X_2;\tilde{Y})\ge \mathbf{c}^\star)
\\&\quad+M_1\Pr(\mathbf{i}(X_1,\tilde{X}_2;\tilde{Y})\ge \mathbf{c}^\star)
\\&\quad+M_2\Pr(\mathbf{i}(\tilde{X}_1,X_2;\tilde{Y})\ge \mathbf{c}^\star))
\\&\le \Pr\left(\mathbf{i}(\tilde{X}_1,\tilde{X}_2;\tilde{Y})\not\ge\mathbf{c}^\star\right)
\\&\quad+M_1M_2K\Pr(i(X_1,X_2;\tilde{Y})\ge c_{12}^\star)
\\&\quad+M_1\Pr(i(X_1;\tilde{Y}|\tilde{X}_2)\ge c_1^\star)
\\&\quad+M_2\Pr(i(X_2;\tilde{Y}|\tilde{X}_1)\ge c_2^\star)
\end{align*}

Note that
\begin{align*}
&\Pr(i(X_1;\tilde{Y}|\tilde{X}_2)\ge c_1^\star)
\\&=\sum_{x_1,x_2,y}p_{X_1}(x_1)p_{\tilde{X}_2,\tilde{Y}}(x_2,y)1\left(i(x_1;y|x_2)\ge c_1^\star\right)
\\&=\sum_{x_1,x_2,y}p_{X_1|X_2}(x_1|x_2)p_{\tilde{X}_2,\tilde{Y}}(x_2,y)\\&\quad\cdot 1\left(i(x_1;y|x_2)\ge c_1^\star\right)
\\&=\sum_{x_2,y}p_{\tilde{X}_2,\tilde{Y}}(x_2,y)
\sum_{x_1}p_{X_1|X_2,Y}(x_1|x_2,y)
\\&\quad\cdot\frac{p_{X_1|X_2}(x_1|x_2)}{p_{X_1|X_2,Y}(x_1|x_2,y)} 
1\left(i(x_1;y|x_2)\ge c_1^\star\right)
\\&=\sum_{x_2,y}p_{\tilde{X}_2,\tilde{Y}}(x_2,y)
\sum_{x_1}p_{X_1|X_2,Y}(x_1|x_2,y)
\\&\quad\cdot\frac{p_{Y|X_2}(y|x_2)}{p_{Y|X_1,X_2}(y|x_1,x_2)} 
\\&\quad\cdot 1\left(\log\frac{p_{Y|X_1,X_2}(y|x_1,x_2)}{p_{Y|X_2}(y|x_2)}\ge c_1^\star\right)
\\&\le \sum_{x_2,y}p_{\tilde{X}_2,\tilde{Y}}(x_2,y)
\sum_{x_1}p_{X_1|X_2,Y}(x_1|x_2,y)\exp(-c_1^\star)
\\&=\exp(-c_1^\star).
\end{align*}
Applying similar arguments to the other terms, we find
\begin{align}
E[P_e]&\le \Pr\left(\mathbf{i}(\tilde{X}_1,\tilde{X}_2;\tilde{Y})\not\ge\mathbf{c}^\star\right)
+M_1M_2K\exp(-c_{12}^\star)\notag
\\&\quad+M_1 \exp(-c_1^\star)
+M_2\exp(-c_2^\star).\label{FBL}
\end{align}

Note that \eqref{FBL} may be viewed as a finite blocklength achievable result. While our primary goal is asymptotic second-order analysis, we proceed by analyzing this bound on the $n$-length product channel. Specifically, we now focus on the case where
$(\cX_1\times\cX_2,p_{Y|X_1,X_2},\cY)$ captures $n$ uses 
of a discrete, memoryless channel.  
We designate this special case by 
\[
(\cX_1^n\times\cX_2^n,(p_{Y|X_1,X_2})^n,\cY^n)
\]
and add superscript $n$ to all coding functions as a reminder of the scenario in operation. Assume that each codeword entry is drawn i.i.d. from $p_{X_1}$ or $p_{X_2}$. Define the CF's score function as
\[
s(x_1^n,x_2^n)=\sum_{i=1}^n i(x_{1i},x_{2i}).
\]
If we choose
\begin{align*}
c_{12}^\star&=\log (M_1M_2K)+\frac{1}{2}\log n,\\
c_1^\star&=\log M_1+\frac{1}{2}\log n,\\
c_2^\star&=\log M_2+\frac{1}{2}\log n,
\end{align*}
then
\begin{align}
&E[P_e]\le \Pr\left(\mathbf{i}(\tilde{X}_1^n,\tilde{X}_2^n;\tilde{Y}^n)\not\ge\mathbf{c}^\star\right)+\frac{3}{\sqrt{n}}\nonumber
\\&\le \Pr\left(i(\tilde{X}_1^n,\tilde{X}_2^n;\tilde{Y}^n)<\log (M_1M_2K)+\frac{1}{2}\log n\right)\nonumber
\\&\quad+\Pr\left(i(\tilde{X}_1^n;\tilde{Y}^n|\tilde{X}_2^n)<\log M_1+\frac{1}{2}\log n\right)\nonumber
\\&\quad+\Pr\left(i(\tilde{X}_2^n;\tilde{Y}^n|\tilde{X}_1^n)<\log M_2+\frac{1}{2}\log n\right)+\frac{3}{\sqrt{n}}.\label{three_terms}
\end{align}

We begin by bounding the second and third terms of \eqref{three_terms} before returning to bound the first. For the second term in \eqref{three_terms}, recall that $\tilde{X}_1^n,\tilde{X}_2^n$ are drawn from the distribution of $(X_1^n,X_2^n)(1,1)$ induced by the cooperation facilitator,
\begin{align}
&\Pr\left(i(\tilde{X}_1^n;\tilde{Y}^n|\tilde{X}_2^n)<\log M_1+\frac{1}{2}\log n\right)\label{hoeffding}
\\&\le K \Pr\left(i(X_1^n;Y^n|X_2^n)<\log M_1+\frac{1}{2}\log n\right)\nonumber
\\&\le K \exp\left\{-\frac{a_1}{n} \left(nI(X_1;Y|X_2)-\log M_1-\frac{1}{2}\log n\right)^2\right\}\nonumber
\end{align}
where the last inequality follows from Hoeffding's inequality and the assumption that $i(X_1;Y|X_2)$ is bounded, where $a_1$ is a constant employed in this bound. By the assumption of $\log M_1$ in the statement of the theorem, this quantity is at most $1/\sqrt{n}$ for a suitable constant $c$.
A similar bound can be applied to the third term in \eqref{three_terms}.

Now we consider the first term in \eqref{three_terms}. For fixed $x_1^n,x_2^n$,
\[
E[i(x_1^n,x_2^n;Y^n)]=\sum_{i=1}^n i(x_{1i},x_{2i})=s(x_1^n,x_2^n).
\]
Thus we can apply the Berry-Esseen theorem to write
\begin{align*}
&\Pr\bigg(i(x_1^n,x_2^n;Y^n)<c_{12}^\star\bigg|(X_1^n,X_2^n)=(x_1^n,x_2^n)\bigg)
\\&\le \Pr\bigg(s(x_1^n,x_2^n)
+\sqrt{\sum_{i=1}^n V(p(y|x_{1i},x_{2i})\|p_Y)}\,Z_0
\\&\qquad<c_{12}^\star\bigg)
+\frac{B}{\sqrt{n}}
\end{align*}
where, as in the statement of the theorem, $Z_0$ is a standard Gaussian random variable.

Assume
\[
V(p_{Y|X_1=x_1,X_2=x_2}\|p_Y)\le V_{\max}\text{ for all }x_1,x_2.
\]
Let
\[
\gamma=V_{\max}\sqrt{\frac{\ln K+\frac{1}{2}\ln n}{2n}}.
\]
Note that
\[
\gamma=O\left(\sqrt{\frac{\log K}{n}}\right)+O\left(\sqrt{\frac{\log n}{n}}\right).
\]
By the assumption that $\log K=o(n)$, $\gamma=o(1)$. By Hoeffding's inequality, we may write
\begin{align*}
    &\Pr\left(\left|\frac{1}{n}\sum_{i=1}^n V(p(y|X_{1i},X_{2i})\|p_Y)-V_2\right|>\gamma\right)
    \\&\le 2\exp\left\{-\frac{2n \gamma^2}{V_{\max}^2}\right\}=\frac{2}{K\sqrt{n}}.
\end{align*}
Thus, by the union bound
\begin{multline*}
\Pr\Bigg(\Bigg|\frac{1}{n}\sum_{i=1}^n V(p(y|f_{1i}(1,k),f_{2i}(1,k))\|p_Y)-V_2\Bigg|>\gamma
\\ \text { for any }k\in[K]\Bigg)\le \frac{2}{\sqrt{n}}.
\end{multline*}
Thus
\begin{align*}
    &E[P_e]\le \Pr\Bigg(s(\tilde{X}_1^n,\tilde{X}_2^n)
    \\&\quad+\sqrt{\sum_{i=1}^n V(p(y|\tilde{X}_{1i},\tilde{X}_{2i})\|p_Y)}\,Z_0<c_{12}^\star\Bigg)
    +O\left(\frac{1}{\sqrt{n}}\right)
    \\&\le E\Bigg[\max_{V'\in[V_2-\gamma,V_2+\gamma]}\Pr\Bigg(s(\tilde{X}_1^n,\tilde{X}_2^n)
    +\sqrt{nV'}\,Z_0
    \\&
    \qquad
    <c_{12}^\star\Bigg|\tilde{X}_1^n,\tilde{X}_2^n\Bigg)\Bigg]
    +O\left(\frac{1}{\sqrt{n}}\right)
    \\&=E\Bigg[\max_{V'\in[V_2-\gamma,V_2+\gamma]}\Phi\left(\frac{c_{12}^\star-s(\tilde{X}_1^n,\tilde{X}_2^n)}{\sqrt{nV'}}\right)\Bigg]
    \\&\quad+O\left(\frac{1}{\sqrt{n}}\right)
\end{align*}
where $\Phi(\cdot)$ is the Gaussian CDF. Similarly, let $\phi(\cdot)$ be the Gaussian PDF. Given $x_1^n,x_2^n$, let
\[
z=\frac{c_{12}^\star-s(x_1^n,x_2^n)}{\sqrt{n}}.
\]
If $z\ge 0$, then we may bound
\begin{align*}
    &\max_{V'\in[V_2-\gamma,V_2+\gamma]}\Phi\left(\frac{z}{\sqrt{V'}}\right)
    \\&=\Phi\left(\frac{z}{\sqrt{V_2-\gamma}}\right)
    \\&=\Phi\left(\frac{z}{\sqrt{V_2}}\right)+\int_{z/\sqrt{V_2}}^{z/\sqrt{V_2-\gamma}} \phi(y)dy
    \\&\le \Phi\left(\frac{z}{\sqrt{V_2}}\right)+\left(\frac{1}{\sqrt{V_2-\gamma}}
    -\frac{1}{\sqrt{V_2}}\right) z \phi\left(\frac{z}{\sqrt{V_2}}\right)
    \\&\le \Phi\left(\frac{z}{\sqrt{V_2}}\right)+\left(\frac{1}{\sqrt{V_2-\gamma}}
    -\frac{1}{\sqrt{V_2}}\right) \sqrt{\frac{V_2}{2\pi e}}
    \\&= \Phi\left(\frac{z}{\sqrt{V_2}}\right)+\left(\sqrt{\frac{V_2}{V_2-\gamma}}
    -1\right) \frac{1}{\sqrt{2\pi e}}.
\end{align*}
If $z\ge 0$, then we may bound
\begin{align*}
    &\max_{V'\in[V_2-\gamma,V_2+\gamma]}\Phi\left(\frac{z}{\sqrt{V'}}\right)
    \\&=\Phi\left(\frac{z}{\sqrt{V_2+\gamma}}\right)
    \\&=\Phi\left(\frac{z}{\sqrt{V_2}}\right)+\int_{z/\sqrt{V_2}}^{z/\sqrt{V_2+\gamma}} \phi(y)dy
    \\&\le \Phi\left(\frac{z}{\sqrt{V_2}}\right)+\left(\frac{1}{\sqrt{V_2}}-\frac{1}{\sqrt{V_2+\gamma}}
    \right) |z| \phi\left(\frac{z}{\sqrt{V_2+\gamma}}\right)
    \\&\le \Phi\left(\frac{z}{\sqrt{V_2}}\right)+\left(\frac{1}{\sqrt{V_2}}-\frac{1}{\sqrt{V_2+\gamma}}
    \right) \sqrt{\frac{V_2+\gamma}{2\pi e}}
    \\&= \Phi\left(\frac{z}{\sqrt{V_2}}\right)+\left(\sqrt{\frac{V_2+\gamma}{V_2}}-1\right) \frac{1}{\sqrt{2\pi e}}.
\end{align*}
Since $\gamma= o(1)$, combining the above bounds gives
\[
\max_{V'\in[V_2-\gamma,V_2+\gamma]}\Phi\left(\frac{z}{\sqrt{V'}}\right)\le \Phi\left(\frac{z}{\sqrt{V_2}}\right)+O(\gamma).
\]
Thus, 
\begin{align*}
&E[P_e]\le E\left[\Phi\left(\frac{c_{12}^\star-s(\tilde{X}_1^n,\tilde{X}_2^n)}{\sqrt{nV_2}}\right)\right]+O(\gamma)+O\left(\frac{1}{\sqrt{n}}\right)
\\&= \Pr\bigg(s(\tilde{X}_1^n,\tilde{X}_2^n)
+\sqrt{nV_2}\,Z_0<c_{12}^\star\bigg)
\\&\qquad+O(\gamma)+O\left(\frac{1}{\sqrt{n}}\right)
\\&= \Pr\bigg(\max_{k\in [K]} \sum_{i=1}^n i(X_{1i}(k),X_{2i}(k))
+\sqrt{nV_2}\,Z_0<c_{12}^\star\bigg)
\\&\qquad+O\left(\sqrt{\frac{\log K}{n}}\right)+O\left(\sqrt{\frac{\log n}{n}}\right).
\end{align*}

\section{Proof of Theorem~\ref{thm:small_deviations}}

Given $\eps$, our goal is to choose $M_1,M_2$ to satisfy the conditions of Theorem~\ref{thm:initial_bound}, while
\begin{equation}\label{M1M2goal}
\log(M_1M_2)=nC_{\text{sum}}
+\sqrt{n}\,F_{S_K}^{-1}(\eps)
-n\theta_n
\end{equation}
where $\theta_n$ satisfies one of \eqref{theta1}--\eqref{theta4} depending on $K$. Given $p_{X_1}p_{X_2}\in\cP^\star$, let $r_1,r_2$ be rates where
\begin{align}
r_1+r_2&=I(X_1,X_2;Y)=C_{\text{sum}},\label{eq:r12_cond}\\
r_1&<I(X_1;Y|X_2),\\
r_2&<I(X_2;Y|X_1).\label{eq:r2_cond}
\end{align}
Let
\[
\log M_j=nr_j+\frac{1}{2}\left[\sqrt{n}\,F_{S_K}^{-1}(\eps)-n\theta_n\right].
\]
This choice clearly satisfies \eqref{M1M2goal}. By Lemma~\ref{lem:max_gausians}, $F_{S_K}^{-1}(\eps)= \sqrt{2V_1\ln K}+O(1)$, so for sufficiently large $n$, \eqref{M1_condition}--\eqref{M2_condition} are easily satisfied. It remains to prove \eqref{eps_condition}. Let $p_e$ be the probability in \eqref{eps_condition}. We divide the remainder of the proof into two cases.

\emph{Case 1:} $K\le \log^{3/2}n$. 
We adopt the notation from the proof of Theorem~\ref{thm:initial_bound}, specifically
\begin{align*}
s(X_1^n,X_2^n)&=\sum_{i=1}^n i(X_{1i}(k),X_{2i}(k)),\\
c_{12}^\star&=\log(M_1M_2K)+\frac{1}{2}\log n.
\end{align*}
Thus
\begin{align*}
    p_e&\le \int_{-\infty}^{\infty} \phi(z) \Pr\bigg(\max_{k\in[K]} s(X_1^n(k),X_2^n(k))
    \\&\qquad<c_{12}^\star-\sqrt{nV_2}\,z\bigg)dz
    \\&=\int_{-\infty}^{\infty} \phi(z) \Pr\bigg(s(X_1^n,X_2^n)<c_{12}^\star-\sqrt{nV_2}\,z\bigg)^K dz.
\end{align*}
Note that $s(X_1^n,X_2^n)$ is an i.i.d.\ sum where each term has expectation 
\begin{equation*}
E[i(X_1,X_2)]
=I(X_1,X_2;Y)=C_{\text{sum}}
\end{equation*}
and variance $V_1$. Thus, by the Berry-Esseen theorem,
\begin{align*}
    p_e&\le \int_{-\infty}^{\infty} \phi(z) \bigg[\Pr\bigg(nC_{\text{sum}}+\sqrt{n}\,\sigma Z_1<c_{12}^\star-\sqrt{nV_2}\,z\bigg)
    \\&\qquad+\frac{B_1}{\sqrt{n}}\bigg]^K dz
\end{align*}
where $Z_1\sim\mathcal{N}(0,1)$.
For any $p\in[0,1]$ and any $0\le q\le 1/K$, we can bound
\begin{align*}
    (p+q)^K&=\sum_{\ell=0}^K \binom{K}{\ell} p^{K-\ell} q^{\ell}
        \\&\le p^K+\sum_{\ell=1}^K \binom{K}{\ell} q^{\ell}
        \\&=p^K+(1+q)^K-1
        \\&\le p^K+e^{qK}-1
        \\&\le p^K+2qK.
\end{align*}
By the assumption that $K\le \log^{3/2}n$, for sufficiently large $n$, $\frac{B_1}{\sqrt{n}}\le \frac{1}{K}$. Thus
\begin{align*}
    &p_e\le \int_{-\infty}^{\infty} \phi(z) \Pr\bigg(nC_{\text{sum}}+\sqrt{n}\,\sigma Z_1<c_{12}^\star-\sqrt{nV_2}\,z\bigg)^K \! dz
    \\&\qquad+ \frac{2B_1 K}{\sqrt{n}}
    \\&=\Pr(nC_{\text{sum}}+\sqrt{n}\,S_K<c_{12}^\star)+O\left(\frac{K}{\sqrt{n}}\right)
    \\&=F_{S_K}\left(\frac{\log(M_1M_2K)+\frac{1}{2}\log n-nC_{\text{sum}}}{\sqrt{n}}\right)+O\left(\frac{K}{\sqrt{n}}\right).
\end{align*}
Recalling Theorem~\ref{thm:initial_bound}, we can achieve probability of error $\eps$ if
\[
p_e+O\left(\sqrt{\frac{\log n}{n}}\right)+O\left(\sqrt{\frac{\log K}{n}}\right)\le\eps.
\]
This condition is satisfied if
\begin{multline*}
\log(M_1M_2K)+\frac{1}{2}\log n
=nC_{\text{sum}}
\\
+\sqrt{n}\,F_{S_K}^{-1}\left(\eps-c_1\frac{K}{\sqrt{n}}
-c_2\sqrt{\frac{\log n}{n}}-c_3\sqrt{\frac{\log K}{n}}
\right)
\end{multline*}
for suitable constants $c_1,c_2,c_3$ and sufficiently large $n$. To simplify the second term, we need the following lemma, which is proved in Appendix~\ref{appendix:inv_cdf_derivative}.
\begin{lemma}\label{lemma:inv_cdf_derivative}
Fix $\eps\in(0,1)$ and $V_1,V_2> 0$. Then
\[
\sup_{K\ge 1}\frac{d}{dp}F_{S_K}^{-1}(p)\bigg|_{p=\eps}<\infty.
\]
\end{lemma}

Applying Lemma~\ref{lemma:inv_cdf_derivative}, there exists a sequence of codes if
\begin{align*}
    \frac{\log(M_1M_2)}{n}
    &\ge C_{\text{sum}}+\frac{1}{\sqrt{n}}\,F_{S_K}^{-1}(\eps)-O\left(\frac{K}{n}\right)
    \\&\qquad-O\left(\frac{\log n}{n}\right).
\end{align*}
This achieves the ranges of $K$ given by \eqref{theta1}--\eqref{theta2}.

\emph{Case 2:} $K\ge\log^{3/2}n$ and $\log K=o(n^{1/3})$. 
For convenience, define
\[
A=\frac{c_{12}^\star-\max_{k\in[K]}s(X_1^n(k),X_2^n(k))}{\sqrt{nV_2}}.
\]
Thus,
\begin{align*}
&p_e =\Pr(Z_0<A)
\\&\le\Pr(Z_0<A,|Z_0|<\sqrt{\ln n})+\Pr(|Z_0|\ge \sqrt{\ln n})
\\&\le\Pr(Z_0<A,|Z_0|<\sqrt{\ln n})+O\left(\frac{1}{\sqrt{n}}\right)
\\&=\int_{-\sqrt{\ln n}}^{\sqrt{\ln n}} \phi(z)\Pr(z<A)dz+O\left(\frac{1}{\sqrt{n}}\right)
\\&=\int_{-\sqrt{\ln n}}^{\sqrt{\ln n}} \phi(z)\Pr\bigg(\max_{k\in[K]}s(X_1^n(k),X_2^n(k))
\\&\qquad<c_{12}^\star-\sqrt{nV_2}\,z\bigg)dz
+O\left(\frac{1}{\sqrt{n}}\right)
\\&=\int_{-\sqrt{\ln n}}^{\sqrt{\ln n}} \phi(z)\Pr\left(s(X_1^n,X_2^n)<c_{12}^\star-\sqrt{nV_2}\,z\right)^K dz
\\&\qquad+O\left(\frac{1}{\sqrt{n}}\right).
\end{align*}
To continue, we need the moderate deviations bound given by the following lemma.

\begin{lemma}[Moderate deviations \cite{chen2013stein}]
Let $X_1,X_2,\ldots$ be i.i.d. random variables with zero mean and unit variance, and let $W=\sum_{i=1}^n X_i/\sqrt{n}$ where $c=E[e^{t|X_1|}]<\infty$ for some $t>0$. There exist constants $a_0$ and $b_0$ depending only on $t$ and $c$ such that, for any $0\le w\le a_0 n^{1/6}$, 
\[
\left|\frac{\Pr(W\ge w)}{Q(w)}-1\right|\le \frac{b_0(1+w^3)}{\sqrt{n}},
\]
where $Q(w)=1-\Phi(w)$ is the complementary CDF of the standard Gaussian distribution.
\end{lemma}

To apply the moderate deviations bound, we can write
\begin{align*}
&\Pr\left(s(X_1^n,X_2^n)<c_{12}^\star-\sqrt{nV_2}\,z\right)
\\&=\Pr\left(\frac{s(X_1^n,X_2^n)-nC_{\text{sum}}}{\sqrt{nV_1}}<w_z\right)
\end{align*}
where
\[
w_z=\frac{c_{12}^\star-\sqrt{nV_2}\,z-nC_{\text{sum}}}{\sqrt{nV_1}}.
\]
Since in our integral, $|z|\le \sqrt{\ln n}$, in order to apply the moderate deviations bound, we need to prove that $|w_z|\le a_0 n^{1/6}$ as long as $|z|\le\sqrt{\ln n}$. We have
\begin{align*}
|w_z|&\le \frac{|c_{12}^\star-nC_{\text{sum}}|}{\sqrt{nV_1}}+\sqrt{\frac{2V_2\ln n}{V_1}}.
\end{align*}
From the target for $M_1M_2$ in \eqref{M1M2goal}, 
\begin{align*}
c_{12}^\star&=\log(M_1M_2K)+\frac{1}{2}\log n
\\&=nC_{\text{sum}}+\sqrt{n}\,F_{S_K}^{-1}(\eps)-n\theta_n+\log K+\frac{1}{2}\log n
\\&\le nC_{\text{sum}}+\sqrt{2V_1n\ln K}+\log K+O(\log n)
\\&=nC_{\text{sum}}+O(n^{2/3}).
\end{align*}
By the assumption that $\log K=o(n^{1/3})$, $\sqrt{n \ln K}\gg \log K$, so
\[
|w_z|=O(\sqrt{\log K})+O(\sqrt{\log n}).
\]
Thus $|w_z|= o(n^{1/6})$, so indeed we may apply the moderate deviations bound.
Let
\begin{align*}
\lambda_n&=\max_{|z|\le\sqrt{\ln n}} \frac{b_0}{\sqrt{n}}(1+|w_z|^3)
\\&=O\left(\frac{\log^{3/2}K}{\sqrt{n}}\right)+O\left(\frac{\log^{3/2}n}{\sqrt{n}}\right)
\end{align*}
 Letting $Z_1\sim\mathcal{N}(0,1)$ we now have
\begin{align*}
    p_e&\le \int_{-\sqrt{\ln{n}}}^{\sqrt{\ln{n}}} \phi(z)\left(1-
\Pr(Z_1>w_z)(1-\lambda_n)\right)^K dz
\\&\qquad+O\left(\frac{1}{\sqrt{n}}\right)
\\&\le \int_{-\sqrt{\ln{n}}}^{\sqrt{\ln{n}}} \phi(z)\left(1-
Q(w_z)(1-\lambda_n)\right)^K dz+O\left(\frac{1}{\sqrt{n}}\right).
\end{align*}
We now claim that for any $w\ge 0$ and any $0\le\lambda\le 3/4$,
\[
Q(w)(1-\lambda)\ge Q(w+2\lambda).
\]
Indeed, it is easy to see that
\[
\frac{Q(w+2\lambda)}{Q(w)}\le \frac{Q(2\lambda)}{Q(0)}=2Q(2\lambda)\le 1-\lambda,
\]
where the last inequality holds if $\lambda\le 3/4$. Note that $\lambda_n=o(1)$, so this inequality holds for sufficiently large $n$.
Thus,
\begin{align*}
p_e
&\le 
E\Bigg[\left(1-Q\left(w_{Z_0}\right)(1-\lambda_n)\right)^K 
 \cdot 1\left(w_{Z_0}\ge 0\right)\Bigg]
\\&\quad+\Pr\left(w_{Z_0}<0\right)+O\left(\frac{1}{\sqrt{n}}\right)
\\&\le E\left[\left(1-Q\left(w_{Z_0}+2\lambda_n\right) \right)^K\right]
+Q\left(\frac{c_{12}^\star-nC_{\text{sum}}}{\sqrt{nV_2}}\right)
\\&\qquad+O\left(\frac{1}{\sqrt{n}}\right).
\end{align*}
Note that
\begin{align*}
    &E\left[\left(1-Q\left(w_{Z_0}+2\lambda_n\right) \right)^K\right]
    \\&=E\left[\Pr\bigg(Z_1<\frac{c_{12}^\star-\sqrt{nV_2}\,Z_0-nC_{\text{sum}}}{\sqrt{nV_1}}+2\lambda_n\bigg|Z_0\bigg)^K\right]
    \\&=\Pr\bigg(\max_{k\in[K]}Z_k<\frac{c_{12}^\star-\sqrt{nV_2}\,Z_0-nC_{\text{sum}}}{\sqrt{nV_1}}+2\lambda_n\bigg)
    \\&=F_{S_K}\left(\frac{c_{12}^\star-nC_{\text{sum}}}{\sqrt{n}}+2\sqrt{V_1}\,\lambda_n\right).
\end{align*}
At this point, we make the choice of $M_1M_2$ slightly more precise; in particular, let
\begin{align*}
    &\log(M_1M_2)=nC_{\text{sum}}
    \\&\quad+\sqrt{n}F_{S_K}^{-1}\left(\eps-c_1\sqrt{\frac{\log K}{n}}-c_2\sqrt{\frac{\log n}{n}}-K^{-V_1/(2V_2)}\right)
    \\&\quad-2\sqrt{nV_1}\,\lambda_n-\frac{1}{2}\log n-\log K
\end{align*}
for suitable constants $c_1$ and $c_2$. From Lemma~\ref{lem:max_gausians},
\begin{align*}
\frac{c_{12}^\star-nC_{\text{sum}}}{\sqrt{n}}
&\ge \sqrt{2V_1\ln K}-o(1).
\end{align*}
Thus
\begin{align*}
    Q\left(\frac{c_{12}^\star-nC_{\text{sum}}}{\sqrt{nV_2}}\right)
    &\le \exp\left\{-\frac{1}{2}\left(\sqrt{\frac{2V_1\ln K}{V_2}}-o(1)\right)^2\right\}
    \\&\le K^{-V_1/(2V_2)}
\end{align*}
where the last inequality holds for sufficiently large $n$.
From Theorem~\ref{thm:initial_bound}, there exists a code with probability of error at most
\begin{align*}
    &p_e+O\left(\sqrt{\frac{\log K}{n}}\right)+O\left(\sqrt{\frac{\log n}{n}}\right)
    \\&\le \eps-c_1\sqrt{\frac{\log K}{n}}-c_2\sqrt{\frac{\log n}{n}}
    \\&\qquad+O\left(\frac{1}{\sqrt{n}}\right)+O\left(\sqrt{\frac{\log K}{n}}\right)+O\left(\sqrt{\frac{\log n}{n}}\right)
    \\&\le \eps
\end{align*}
assuming $c_1,c_2$ are chosen properly. This proves that we can achieve the sum-rate
\begin{align*}
&\frac{\log(M_1M_2)}{n}\ge C_{\text{sum}}
\\&\quad+\frac{1}{\sqrt{n}}F_{S_K}^{-1}\left(\eps-c_1\sqrt{\frac{\log K}{n}}-c_2\sqrt{\frac{\log n}{n}}-K^{-V_1/(2V_2)}\right)
    \\&\quad-\frac{2\sqrt{V_1}\,\lambda_n}{\sqrt{n}}-\frac{\log n}{2n}-\frac{\log K}{n}
    \\&\ge C_{\text{sum}}+\frac{1}{\sqrt{n}}F_{S_K}^{-1}(\eps)-O\left(\frac{\log^{3/2} K}{n}\right)
    -O\left(\frac{\log^{3/2} n}{n}\right)
\end{align*}
where in the last inequality we have used Lemma~\ref{lemma:inv_cdf_derivative} as well as the bound on $\lambda_n$. This achives the ranges of $K$ given by \eqref{theta3}--\eqref{theta4}.

\section{Proof of Theorem~\ref{thm:large_deviations}}

This proof uses the method of types. A probability mass function $p_X$ is an $n$-length type on alphabet $\cX$ if $p_X(x)$ is a multiple of $1/n$ for each $x\in\cX$. For an $n$-length type $p_X$, the type class is denoted $T(p_X)$.

Let $p_{X_1,X_2}$ be an $n$-length joint type on alphabet $\cX_1\times\cX_2$. Note that the marginal distributions $p_{X_1}$ and $p_{X_2}$ are also $n$-length types. We employ the following random code construction. Draw codewords uniformly from the type classes $T(p_{X_1})$ and $T(p_{X_2})$. Given message pair $(m_1,m_2)$, the cooperation facilitator chooses uniformly from the set of $k\in[K]$ where
\[
(f(m_1,k),f(m_2,k))\in T(p_{X_1,X_2}).
\]
If there is no such $k$, the CF chooses $k$ uniformly at random. These random choices at the CF are taken to be part of the random code design. For the purposes of this proof, the three information densities employ the joint distribution $p_{X_1,X_2}$. The quantity $V_2$ is also defined as in \eqref{V_def} using information density for this joint distribution. The decoder is as follows. Given $y^n$, choose the unique message pair $(m_1,m_2)$ such that
\begin{enumerate}
\item $\mathbf{i}((X_1^n,X_2^n)(m_1,m_2);y^n)\ge \mathbf{c}^\star$,
\item $((X_1^n,X_2^n)(m_1,m_2))\in T(p_{X_1,X_2})$
\end{enumerate}
for a constant vector $\mathbf{c}^\star=[c_{12}^\star,c_1^\star,c_2^\star]^T$ to be determined. If there is no message pair or more than one satisfying these conditions, declare an error. Note that, given 
\[(X_1^n,X_2^n)(m_1,m_2)\in T(p_{X_1,X_2}),\]
$(X_1^n,X_2^n)(m_1,m_2)$ is uniformly distributed on $T(p_{X_1,X_2})$. Let $q(x_1^n,x_2^n)$ be the uniform distribution on the type class $T(p_{X_1,X_2})$, with corresponding conditional distributions $q(x_1^n|x_2^n)$ and $q(x_2^n|x_1^n)$. Define random variables $X_1^n,X_2^n,Y^n$ to have distribution
\[
p_{X_1^n,X_2^n,Y^n}(x_1^n,x_2^n,y^n)=q(x_1^n,x_2^n) p_{Y^n|X_1^n,X_2^n}(y^n|x_1^n,x_2^n).
\]
Furthermore, define $Y_1^n,Y_2^n,Y_{12}^n$ where
\begin{multline*}
p_{Y_1^n,Y_2^n,Y_{12}^n|X_1^n,X_2^n,Y^n}(y_1^n,y_2^n,y_{12}^n|x_1^n,x_2^n,y^n)
\\=p_{Y^n|X_2^n}(y_1^n|x_2^n)p_{Y^n|X_1^n}(y_2^n|x_1^n) p_{Y^n}(y_{12}^n)
\end{multline*}
Now we may bound the expected error probability by
\begin{align}
&E[P_e]
\\&\le \Pr((X_1^n,X_2^n)(1,1)\notin T(p_{X_1,X_2}))\notag
\\&\quad+\Pr(\mathbf{i}((X_1^n,X_2^n)(1,1);Y^n)\not\ge \mathbf{c}^\star)\notag
\\&\quad
+\sum_{(\hat{m}_1,\hat{m}_2)\ne(1,1)}\Pr\bigg((X_1^n,X_2^n)(\hat{m}_1,\hat{m}_2)\in T(p_{X_1,X_2}),\notag
\\&\quad\mathbf{i}((X_1^n,X_2^n)(\hat{m}_1,\hat{m}_2);Y^n)\ge \mathbf{c}^\star)\bigg|(X_1^n,X_2^n)(1,1)\bigg)\notag
\\&\le \Pr((X_1^n,X_2^n)(1,1)\notin T(p_{X_1,X_2}))\notag
\\&\quad+\Pr(\mathbf{i}((X_1^n,X_2^n)(1,1);Y^n)\not\ge \mathbf{c}^\star)\notag
\\&\quad+\sum_{(\hat{m}_1,\hat{m}_2)\ne(1,1)}\Pr\bigg(\mathbf{i}((X_1^n,X_2^n)(\hat{m}_1,\hat{m}_2);Y^n)\ge \mathbf{c}^\star)\notag
\\&\quad\bigg|(X_1^n,X_2^n)(1,1),(X_1^n,X_2^n)(\hat{m}_1,\hat{m}_2)\in T(p_{X_1,X_2})\bigg).\label{error_summation}
\end{align}
In the summation in \eqref{error_summation}, consider a term where $\hat{m}_1\ne 1$ and $\hat{m}_2\ne 1$. In this case, $(X_1^n,X_2^n)(\hat{m}_1,\hat{m}_2)$ is independent from $Y^n$, so we may write that
\[
((X_1^n,X_2^n)(\hat{m}_1,\hat{m}_2),Y^n)\stackrel{d}{=}(X_1^n,X_2^n,Y_{12}^n),
\]
where $Y_{12}^n$ has the same distribution as $Y^n$ but is independent from $X_1^n,X_2^n$; i.e.,
\[
p_{Y_{12}^n|X_1^n,X_2^n}(y^n|x_1^n,x_2^n)=p_{Y^n}(y^n).
\]
Now consider a term in \eqref{error_summation} where $\hat{m}_1=1$ but $\hat{m}_2\ne 1$. In this case, whether the transmitted signal from user 1 with message pair $(1,\hat{m}_2)$ is the same as that with message pair $(1,1)$ depends on whether $e(1,\hat{m}_2)=e(1,1)$. Thus, the term in \eqref{error_summation} is no more than
\begin{align*}
&\Pr\bigg(\mathbf{i}((X_1^n,X_2^n)(1,\hat{m}_2);Y^n)\ge \mathbf{c}^\star)\notag
\bigg|(X_1^n,X_2^n)(1,1),
\\&\quad(X_1^n,X_2^n)(1,\hat{m}_2)\in T(p_{X_1,X_2}),e(1,\hat{m}_2)=e(1,1)\bigg)
\\&\quad+\Pr\bigg(\mathbf{i}((X_1^n,X_2^n)(1,\hat{m}_2);Y^n)\ge \mathbf{c}^\star)\notag
\bigg|(X_1^n,X_2^n)(1,1),
\\&\quad(X_1^n,X_2^n)(1,\hat{m}_2)\in T(p_{X_1,X_2}),e(1,\hat{m}_2)\ne e(1,1)\bigg).
\end{align*}
In the first term, $Y^n$ is the channel output where $X_1^n(1,\hat{m}_2)$ is one of the channel inputs, but the channel input for user 2 is unrelated. However, by the condition that $(X_1^n,X_2^n)(1,\hat{m}_2)\in T(p_{X_1,X_2})$, these two codewords are distributed according to $q(x_1^n,x_2^n)$. Thus we may write that
\[
((X_1^n,X_2^n)(1,\hat{m}_2),Y^n)\stackrel{d}{=}(X_1^n,X_2^n,Y_{2}^n),
\]
where
\[
p_{Y_{2}^n|X_1^n,X_2^n}(y^n|x_1^n,x_2^n)=p_{Y^n|X_1^n}(y^n|x_1^n).
\]
In the second term, the transmitted signals are unrelated, and so the three sequences once again have the same distribution as $(X_1^n,X_2^n,Y_{12}^n)$. We may apply a similar analysis for the case where $\hat{m}_1\ne 1$ and $\hat{m}_2=1$, defining $Y_1^n$ by
\[
p_{Y_{1}^n|X_1^n,X_2^n}(y^n|x_1^n,x_2^n)=p_{Y^n|X_2^n}(y^n|x_2^n).
\]
Therefore
\begin{align*}
    E[P_e]
    &\le
    \Pr((X_1^n,X_2^n)(1,1)\notin T(p_{X_1,X_2}))
\\&\quad+\Pr(\mathbf{i}(X_1^n,X_2^n;Y^n)\not\ge \mathbf{c}^\star)
\\&\quad+M_1M_2\Pr(\mathbf{i}(X_1^n,X_2^n;Y_{12}^n)\ge \mathbf{c}^\star)
\\&\quad+M_1\Pr(\mathbf{i}(X_1^n,X_2^n;Y_1^n)\ge \mathbf{c}^\star)
\\&\quad+M_2\Pr(\mathbf{i}(X_1^n,X_2^n;Y_2^n)\ge \mathbf{c}^\star)
\\&\le \Pr((X_1^n,X_2^n)(1,1)\notin T(p_{X_1,X_2}))
\\&\quad+\Pr(\mathbf{i}(X_1^n,X_2^n;Y^n)\not\ge \mathbf{c}^\star)
\\&\quad+M_1M_2\Pr(i(X_1^n,X_2^n;Y_{12}^n)\ge c_{12}^\star)
\\&\quad+M_1\Pr(i(X_1^n;Y_1^n|X_2^n)\ge c_1^\star)
\\&\quad+M_2\Pr(i(X_2^n;Y_2^n|X_1^n)\ge c_2^\star).
\end{align*}

For any $(x_1^n,x_2^n)\in T(p_{X_1,X_2})$,
\begin{align*}
    &q(x_1^n,x_2^n)
    \\&=\frac{1}{|T(p_{X_1,X_2})|}
    \\&\le (n+1)^{|\cX_1|\cdot|\cX_2|}2^{-nH(X_1,X_2)}
    \\&= (n+1)^{|\cX_1|\cdot|\cX_2|}\prod_{x_1,x_2} p_{X_1,X_2}(x_1,x_2)^{np_{X_1,X_2}(x_1,x_2)}
    \\&= (n+1)^{|\cX_1|\cdot|\cX_2|}\prod_{i=1}^n p_{X_1,X_2}(x_{1i},x_{2i}).
\end{align*}
Thus, for any $x_1^n,x_2^n$ including those not in $T(p_{X_1,X_2})$,
\[
q(x_1^n,x_2^n)\le (n+1)^{|\cX_1|\cdot|\cX_2|}\prod_{i=1}^n p_{X_1,X_2}(x_{1i},x_{2i}).
\]
By similar calculations
\begin{align*}
    q(x_1^n|x_2^n)\le (n+1)^{|\cX_1|\cdot|\cX_2|}\prod_{i=1}^n p_{X_1|X_2}(x_{1i}|x_{2i}),
    \\
    q(x_2^n|x_1^n)\le (n+1)^{|\cX_1|\cdot|\cX_2|}\prod_{i=1}^n p_{X_2|X_1}(x_{2i}|x_{1i}).
\end{align*}
We bound $\Pr(i(X_1^n,X_2^n;Y_{12}^n)\ge c_{12}^\star)$ as
\begin{align*}
&\Pr(i(X_1^n,X_2^n;Y_{12}^n)\ge c_{12}^\star)
\\&=\sum_{x_1^n,x_2^n,y^n} q(x_1^n,x_2^n) p_{Y^n}(y^n)1(i(x_1^n,x_2^n;y^n)\ge c_{12}^\star)
\\&\le (n+1)^{|\cX_1|\cdot|\cX_2|}\sum_{x_1^n,x_2^n,y^n} \prod_{i=1}^n p_{X_1,X_2}(x_{1i},x_{2i})
p_{Y^n}(y^n)
\\&\qquad\cdot 1\left(\sum_{i=1}^n i(x_{1i},x_{2i};y_i)\ge c_{12}^\star\right)
\\&\le (n+1)^{|\cX_1|\cdot|\cX_2|}\exp\{-c_{12}^\star\}
\\&\quad\cdot\sum_{x_1^n,x_2^n,y^n} \prod_{i=1}^n p_{X_1,X_2|Y}(x_{1i},x_{2i}|y_i)
p_{Y^n}(y^n)
\\&=(n+1)^{|\cX_1|\cdot|\cX_2|}\exp\{-c_{12}^\star\}.
\end{align*}
Using similar bounds on the other terms, we have
\begin{align*}
    E[P_e]
    &\le \Pr((X_1^n,X_2^n)(1,1)\notin T(p_{X_1,X_2}))
\\&\quad+\Pr(\mathbf{i}(X_1^n,X_2^n;Y^n)\not\ge \mathbf{c}^\star)
\\&\quad+(n+1)^{|\cX_1|\cdot|\cX_2|}\Big(M_1M_2\exp\{-c_{12}^\star\}
\\&\quad+M_1\exp\{-c_1^\star\}+M_2\exp\{-c_2^\star\}\Big).
\end{align*}
Next, choose
\begin{align*}
c_{12}^\star&=\log (M_1M_2)+\frac{1}{2}\log n+|\cX_1|\cdot|\cX_2|\log(n+1),\\
c_1^\star&=\log M_1+\frac{1}{2}\log n+|\cX_1|\cdot|\cX_2|\log(n+1),\\
c_2^\star&=\log M_2+\frac{1}{2}\log n+|\cX_1|\cdot|\cX_2|\log(n+1).
\end{align*}
Then
\begin{align}
    E[P_e]
    &\le \Pr((X_1^n,X_2^n)(1,1)\notin T(p_{X_1,X_2}))\nonumber
    \\&\quad+\Pr(\mathbf{i}(X_1^n,X_2^n;Y^n)\not\ge \mathbf{c}^\star)
    +\frac{3}{\sqrt{n}}\nonumber
    \\&\le \Pr((X_1^n,X_2^n)(1,1)\notin T(p_{X_1,X_2}))\nonumber
    \\&\quad+\Pr(i(X_1^n,X_2^n;Y^n)< c_{12}^\star)\nonumber
    \\&\quad+\Pr(i(X_1^n;Y^n|X_2^n)< c_{1}^\star)\nonumber
    \\&\quad+\Pr(i(X_2^n;Y^n|X_2^n)< c_{2}^\star)+\frac{3}{\sqrt{n}}.\label{four_terms}
\end{align}

As in the proof of Thm.~\ref{thm:small_deviations}, let $(r_1,r_2)$ be a pair of rates satisfying \eqref{eq:r12_cond}--\eqref{eq:r2_cond}.
We now choose
\[
\log M_j=r_j-\frac{1}{2}\left(\sqrt{nV_2}\,Q^{-1}(\eps)+n\theta_n\right),\quad j=1,2
\]
where $\theta_n$ is an error term to be determined chosen below to satisfy $\theta_n\le O(\frac{\log n}{n})$. Thus
\[
\log(M_1M_2)=nI(X_1,X_2;Y)-\sqrt{nV_2}\,Q^{-1}(\eps)-n\theta_n.
\]

Consider the first term in \eqref{four_terms}. Note that $(X_1^n,X_2^n)(1,1)\notin T(p_{X_1,X_2})$ only if 
\[
(f_1(1,k),f_2(1,k))\notin T(p_{X_1,X_2})\text{ for all }k\in[K].
\]
This occurs with probability bounded as
\begin{align*}
&\Pr((X_1^n,X_2^n)(1,1)\notin T(p_{X_1,X_2}))
\\&=\left(1-\frac{|T(p_{X_1,X_2})|}{|T(p_{X_1})|\cdot|T(p_{X_2})|}\right)^K
\\&\le \left(1-(n+1)^{-|\cX_1|\cdot|\cX_2|} 2^{-nI(X_1;X_2)}\right)^K
\\&\le \exp\left\{-K(n+1)^{-|\cX_1|\cdot|\cX_2|} 2^{-nI(X_1;X_2)}\right\}
\\&\le \frac{1}{\sqrt{n}},
\end{align*}
where the last inequality holds if
\begin{align}
I(X_1;X_2)&\le \frac{1}{n}\bigg(\log K-|\cX_1|\cdot|\cX_2|\log(n+1)\nonumber
\\&\quad-\log\left(\frac{1}{2}\ln n\right)\bigg).\label{MI_condition}
\end{align}

Now consider the second term in \eqref{four_terms}. For any $(x_1^n,x_2^n)\in T(p_{X_1,X_2})$,
\begin{align*}
&\sum_{i=1}^n E[i(x_{1i},x_{2i};Y_i)]=nI(X_1,X_2;Y),\\
&\sum_{i=1}^n \text{Var}[i(x_{1i},x_{2i};Y_i)]
\\&\quad=n\sum_{x_1,x_2} p_{X_1,X_2}(x_1,x_2)V(p_{Y|X_1=x_1,X_2=x_2}\|p_Y)=nV_2.
\end{align*}
By the Berry-Esseen inequality, 
\begin{align*}
    &\Pr(i(X_1^n,X_2^n;Y^n)< c_{12}^\star)
    \\&\le \max_{(x_1^n,x_2^n)\in T(p_{X_1,X_2})} \Pr(i(x_1^n,x_2^n;Y^n)< c_{12}^\star|x_1^n,x_2^n)
    \\&\le Q\left(\frac{nI(X_1,X_2;Y)-c_{12}^\star}{\sqrt{nV_2}}\right)+O\left(\frac{1}{\sqrt{n}}\right).
  \end{align*}

As in the proof of  Thm.~\ref{thm:small_deviations} (near \eqref{hoeffding}), we use Hoeffding's inequality to bound the third and fourth terms in  \eqref{four_terms} from above by $1/\sqrt{n}$.

Putting together all the above bounds, for any $p_{X_1,X_2}$ satisfying \eqref{MI_condition}, we find
\begin{align*}
    &E[P_e]
    \\&\le Q\left(\frac{nI(X_1,X_2;Y)-c_{12}^\star}{\sqrt{nV_2}}\right)+O\left(\frac{1}{\sqrt{n}}\right)
    \\&=Q\left(\frac{nI(X_1,X_2;Y)-\log(M_1M_2)-O(\log n)}{\sqrt{nV_2}}\right)
    \\&\qquad+O\left(\frac{1}{\sqrt{n}}\right)
    \\&=Q\left(Q^{-1}(\eps)+\sqrt{\frac{n}{V_2}}\,\theta_n-O\left(\frac{\log n}{\sqrt{n}}\right)\right)+O\left(\frac{1}{\sqrt{n}}\right).
\end{align*}
There exists a choice for $\theta_n= O(\frac{\log n}{n})$ where this bound is no greater than $\eps$. This proves that we can achieve the sum-rate
\begin{align*}
    \frac{\log(M_1M_2)}{n}
    &\ge I(X_1,X_2;Y)-\sqrt{\frac{V_2}{n}}\,Q^{-1}(\eps)-O\left(\frac{\log n}{n}\right)
\end{align*}
for any $p_{X_1,X_2}$ satisfying \eqref{MI_condition}.

\appendices

\section{Proof of Lemma~\ref{large_but_small}}\label{appendix:large_but_small}

Through this proof, $x\approx y$ means that $x-y\to 0$ as $a\to 0$.
For small $a$, $I(X_1;X_2)\le a$ implies that $p_{X_1,X_2}\approx p_{X_1}p_{X_2}$. Thus, the second-order Taylor approximation for the mutual information gives
\begin{multline*}
I(X_1;X_2)\\
\approx\frac{1}{2\ln 2} \sum_{x_1,x_2} \frac{(p_{X_1,X_2}(x_1,x_2)-p_{X_1}(x_1)p_{X_2}(x_1))^2}{p_{X_1}(x_1)p_{X_2}(x_2)}.
\end{multline*}
Moreover, the first-order Taylor approximation of the mutual information $I(X_1,X_2;Y)$ is
\[
\sum_{x_1,x_2,y} p_{X_1,X_2}(x_1,x_2)\log \frac{p_{Y|X_1,X_2}(y|x_1,x_2)}{p_Y(y)}
\]
where
\[
p_Y(y)=\sum_{x_1,x_2}p_{X_1}(x_1)p_{X_2}(x_2)p_{Y|X_1,X_2}(y|x_1,x_2).
\]
As usual, let
\begin{align*}
i(x_1,x_2;y)&=\log \frac{p_{Y|X_1,X_2}(y|x_1,x_2)}{p_Y(y)}
\\ i(x_1,x_2)&=\sum_y p_{Y|X_1,X_2}(y|x_1,x_2) i(x_1,x_2;y).
\end{align*}
Also let
\[
I_0(X_1,X_2;Y)=\sum_{x_1,x_2} p_{X_1}(x_1)p_{X_2}(x_2) i(x_1,x_2)
\]
be the mutual information where $X_1$ and $X_2$ are independent. We can now rewrite the optimization problem for $\Delta(a)$ in terms of the marginal distributions $p_{X_1},p_{X_2},$ and
\[
r(x_1,x_2)=p_{X_1,X_2}(x_1,x_2)-p_{X_1}(x_1)p_{X_2}(x_2).
\]
Note that
\begin{multline*}
I(X_1,X_2;Y)-C_{\text{sum}}\\
\approx \sum_{x_1,x_2} r(x_1,x_2)i(x_1,x_2)+I_0(X_1,X_2;Y)-C_{\text{sum}}
\end{multline*}

In particular, if we consider maximizing over only $r$, the optimization problem is
\begin{equation}\label{r_opt}
\begin{array}{ll}
\text{maximize} &
\sum_{x_1,x_2} r(x_1,x_2) i(x_1,x_2)\\
\text{subject to} & \sum_{x_1,x_2} \frac{r(x_1,x_2)^2}{p_{X_1}(x_1)p_{X_2}(x_2)}\le a\, 2\ln 2
\\&\sum_{x_2} r(x_1,x_2)=0\text{ for all }x_1\in\cX_1.
\\&\sum_{x_1} r(x_1,x_2)=0\text{ for all }x_2\in\cX_2
\end{array}
\end{equation}
The Lagrangian for this problem is
\begin{align*}
    &\sum_{x_1,x_2} r(x_1,x_2) i(x_1,x_2)-\lambda\left(\frac{r(x_1,x_2)^2}{p_{X_1}(x_1)p_{X_2}(x_2)}- a\, 2\ln 2\right)
    \\&\quad+\sum_{x_1}\nu_1(x_1)\sum_{x_2}r(x_1,x_2)
    +\sum_{x_2}\nu_2(x_2)\sum_{x_1}r(x_1,x_2).
\end{align*}
Differentiating with respect to $r(x_1,x_2)$ and setting to zero, we find that  the optimal $r(x_1,x_2)$ is of the form
\[
r(x_1,x_2)=\frac{p_{X_1}(x_1)p_{X_2}(x_2)}{2\lambda}\left(i(x_1,x_2)+\nu_1(x_1)+\nu_2(x_2)\right).
\]
We first find the values of the dual variables $\nu_1$ and $\nu_2$. For any $x_1$, we need
\begin{align*}
    0&=\sum_{x_2}r(x_1,x_2)
    \\&=\frac{p_{X_1}(x_1)}{2\lambda}\left(E[i(x_1,X_2)]+\nu_1(x_1)+E[\nu_2(X_2)]\right)
\end{align*}
where the expectations are with respect to $(X_1,X_2)\sim p_{X_1}p_{X_2}$. Combining this constraint with the equivalent one for $\nu_2$, we  must have
\begin{align*}
\nu(x_1)&=-E[i(x_1,X_2)]-E[\nu_2(X_2)]
\\\nu(x_2)&=-E[i(X_1,x_2)]-E[\nu_1(X_1)].
\end{align*}
Taking the expectation of either constraint gives
\[
E[\nu_2(X_2)]+E[\nu(X_1)]=-E[i(X_1,X_2)].
\]
Thus
\begin{multline*}
\nu_1(x_1)+\nu_2(x_2)\\=-E[i(x_1,X_2)]-E[i(X_1,x_2)]+E[i(X_1,X_2)].
\end{multline*}
and so
\[
r(x_1,x_2)=\frac{1}{2\lambda}p_{X_1}(x_1)p_{X_2}(x_2) j(x_1,x_2)
\]
where
\begin{multline*}
j(x_1,x_2)=i(x_1,x_2)-E[i(x_1,X_2)]
\\-E[i(X_1,x_2)]+E[i(X_1,X_2)].
\end{multline*}
To find $\lambda$, we use the constraint
\begin{align*}
a\,2\ln 2&=\sum_{x_1,x_2} \frac{r(x_1,x_2)^2}{p_{X_1}(x_1)p_{X_2}(x_2)}
\\&=\frac{1}{(2\lambda)^2}E[j(X_1,X_2)^2]
\end{align*}
so
\[
\frac{1}{2\lambda}=\sqrt{\frac{a\,2\ln 2}{E[j(X_1,X_2)^2]}}.
\]
We may now derive the optimal objective value for the optimization problem in \eqref{r_opt}, which is
\begin{multline*}
    \sum_{x_1,x_2}r(x_1,x_2)i(x_1,x_2)
    \\=\sqrt{\frac{a\,2\ln 2}{E[j(X_1,X_2)^2]}} \, E[j(X_1,X_2)i(X_1,X_2)].
\end{multline*}
Now considering the optimization over $p_{X_1},p_{X_2}$, we may write
\begin{align*}
    \Delta(a)&\approx \max_{p_{X_1}p_{X_2}} \sqrt{\frac{a\,2\ln 2}{E[j(X_1,X_2)^2]}} \, E[j(X_1,X_2)i(X_1,X_2)]
    \\&\qquad+I_0(X_1,X_2;Y)-C_{\text{sum}}.
\end{align*}
Note that for small $a$, the RHS will be negative unless $p_{X_1},p_{X_2}$ are such that $I_0(X_1,X_2;Y)=C$ (i.e., they are sum-capacity achieving). By the optimality conditions for the maximization defining the sum-capacity, this implies that 
\begin{align*}
    E[i(x_1,X_2)]&=C_{\text{sum}}\text{ for all }x_1\text{ where }p_{X_1}(x_1)>0\\
    E[i(X_1,x_2)]&=C_{\text{sum}}\text{ for all }x_1\text{ where }p_{X_2}(x_2)>0.
\end{align*}
Thus, for $x_1,x_2$ where $p_{X_1}(x_1)p_{X_2}(x_2)>0$, we have
\[
j(x_1,x_2)=i(x_1,x_2)-C.
\]
Thus $E[j(X_1,X_2)^2]=\text{Var}(i(X_1,X_2))$, and
\begin{align*}
E[j(X_1,X_2)i(X_1,X_2)]&=E[i(X_1,X_2)^2-C_{\text{sum}}\,i(X_1,X_2)]
\\&=E[i(X_1,X_2)^2]-C_{\text{sum}}^2
\\&=\text{Var}(i(X_1,X_2)).
\end{align*}
Therefore
\begin{align*}
    \Delta(a)&\approx\max_{p_{X_1}p_{X_2}:I(X_1,X_2;Y)=C_{\text{sum}}} \sqrt{a\,2\ln 2\,\text{Var}(i(X_1,X_2))}
    \\&=\sigma \sqrt{a\,2\ln 2}.
\end{align*}

\section{Proof of Lemma~\ref{lemma:inv_cdf_derivative}}\label{appendix:inv_cdf_derivative}

We first need the following lemma.

\begin{lemma}\label{lemma:indep_var_pdf}
Fix $\eps\in(0,1)$. Let $Y$ and $Z$ be independent random variables where
\begin{align*}
f_Y(y)\ge c \text{ for all }p\in[F_Y^{-1}(\eps/4),F_Y^{-1}({\textstyle\frac{3+\eps}{4}})],\\
d\ge f_Z(y)\ge c \text{ for all }p\in[F_Z^{-1}(\eps/4),F_Z^{-1}({\textstyle\frac{3+\eps}{4}})].
\end{align*}
Then for $X=Y+Z$,
\[
f_X(F_X^{-1}(\eps))\ge\min\left\{\frac{3c}{4},\frac{c^2\eps}{4d}\right\}.
\]
\end{lemma}
\begin{IEEEproof}
Let $x=F_X^{-1}(\eps)$. Note that
\begin{align*}
    F_X(y+z)&=\Pr(Y+Z\le y+z)
    \\&\le \Pr(Y\le y\text{ or }Z\le z)
    \\&\le F_Y(y)+F_Z(z).
\end{align*}
In particular,
\[
F_X(F_Y^{-1}(\eps/2)+F_Z^{-1}(\eps/2))\le\eps
\]
so
\[
x\ge F_Y^{-1}(\eps/2)+F_Z^{-1}(\eps/2).
\]
By similar reasoning,
\[
x\le F_Y^{-1}\left(\frac{1+\eps}{2}\right)+F_Z^{-1}\left(\frac{1+\eps}{2}\right).
\]
Define
\begin{align*}
    y_1&=F_Y^{-1}(\eps/4), & y_2&=F_Y^{-1}({\textstyle\frac{3+\eps}{4}}),\\
    z_1&=F_Z^{-1}(\eps/4), & z_2&=F_Z^{-1}({\textstyle\frac{3+\eps}{4}}).
\end{align*}
Consider several cases. First, if 
\begin{equation}\label{xyz_cond1}
y_2+z_1\le x\le y_1+z_2.
\end{equation}
Then
\begin{align*}
    f_X(x)&=\int_{-\infty}^\infty f_Y(x-z) f_Z(z)dz
    \\&\ge \int_{z_1}^{z_2}c f_Y(x-z)dz
    \\&=c \Pr(x-z_2<Y<x-z_1)
    \\&\ge c\Pr(y_1<Y<y_2)
    \\&=c\left(\frac{3+\eps}{4}-\frac{\eps}{4}\right)
    \\&\ge \frac{3c}{4}.
\end{align*}
Similarly, if 
\begin{equation}\label{xyz_cond2}
y_1+z_2\le x\le y_2+z_1,
\end{equation}
then $f_X(x)\ge \frac{3c}{4}$. Now consider the case that neither \eqref{xyz_cond1} nor \eqref{xyz_cond2} holds. We have
\begin{align*}
    f_X(x)&\ge \int_{\max\{y_1,x-z_2\}}^{\min\{y_2,x-z_1\}} c^2 dy
    \\&=c^2\left[\min\{y_2,x-z_1\}-\max\{y_1,x-z_2\}\right]
\end{align*}
By the assumption that \eqref{xyz_cond1} and \eqref{xyz_cond2} do not hold, we have
\[
f_X(x)\ge c^2\min\{y_2+z_2-x,x-y_1-z_2\}.
\]
Note that
\begin{align*}
    y_2+z_2-x&\ge F_Y^{-1}(1/2+\eps)+F_Z^{-1}(1/2+\eps)
    \\&\qquad-F_Y^{-1}\left(\frac{1+\eps}{2}\right)+F_Z^{-1}\left(\frac{1+\eps}{2}\right)
    \\&\ge F_Z^{-1}(1/2+\eps)-F_Z^{-1}\left(\frac{1+\eps}{2}\right)
    \\&\ge \frac{\eps}{2d}.
\end{align*}
Moreover
\begin{align*}
    x-y_1-y_2&\ge F_Y^{-1}({\textstyle\frac{\eps}{2}})+F_Z^{-1}({\textstyle\frac{\eps}{2}})
    -F_Y^{-1}({\textstyle\frac{\eps}{4}})-F_Z^{-1}({\textstyle\frac{\eps}{4}})
    \\&\ge F_Z^{-1}(\eps/2)-F_Z^{-1}(\eps/4)
    \\&\ge \frac{\eps}{4d}.
\end{align*}
Thus in this case,
\[
f_X(x)\ge \frac{c^2 \eps}{2d}.
\]
\end{IEEEproof}

We now complete the proof of Lemma~\ref{lemma:inv_cdf_derivative}.
Recall that $S_K=\sqrt{V_1}Z(K)+\sqrt{V_2}Z_0$ where $Z(K)=\max_{k\in[K]}Z_k$. Let $x=F_{S_K}^{-1}(\eps)$. Note that
\[
\frac{d}{ds}F_{S_K}(s)=f_{S_K}(s)
\]
and so
\[
\frac{d}{dp}F_{S_K}^{-1}(p)\bigg|_{p=\eps}=\frac{1}{f_{S_K}(x)}.
\]
Thus it is sufficient to show that $f_{S_K}(x)$ is bounded away from zero for all $K$. Since $Z_0\sim\mathcal{N}(0,1)$, 
\[
F_{Z_0}^{-1}(\eps/4)\!\ge\! -\sqrt{2\ln(2/\eps)},
\quad F_{Z_0}^{-1}({\textstyle\frac{3+\eps}{4}})\!\le\! \sqrt{2\ln(2/(1-\eps))}.
\]
Thus, for $z\in[F_{Z_0}^{-1}(\eps/4),F_{Z_0}^{-1}({\textstyle\frac{3+\eps}{4}})]$,
\begin{align*}
    f_{Z_0}(z)&=\phi(z)
    \\&\ge \max\{\phi(-\sqrt{2\ln(2/\eps)}),\phi(\sqrt{2\ln(2/(1-\eps))})\}
    \\&= \max\left\{\frac{2}{\sqrt{2\pi}\eps},\frac{2}{\sqrt{2\pi}(1-\eps)}\right\}
    \\&=\frac{2}{\sqrt{2\pi}\min\{\eps,1-\eps\}}.
\end{align*}
Moreover, for all $z$,
\[
f_{Z_0}(z)\le \frac{1}{\sqrt{2\pi}}.
\]

Now we prove a lower bound on $f_{Z(K)}(y)$. Specifically let $y=F_{Z(K)}^{-1}(p)$ for $p\in[\frac{\eps}{4},\frac{3+\eps}{4}]$. Note that
\[
p=F_{Z(K)}(y)=\Phi(y)^K
\]
so $\Phi(y)=p^{1/K}$. We have
\begin{align*}
    f_{Z(K)}(y)&=K\Phi(y)^{K-1}\phi(y)
    \\&=K p^{1-1/K}\phi(y)
\end{align*}
Suppose $p^{1/K}<1/2$. Thus $y<0$. Also, since $K\ge 1$, $p^{1/K}\ge p\ge \frac{\eps}{4}$, we have
\begin{align*}
    \frac{\eps}{4}&\le p^{1/K}
    \\&=\Phi(y)
    \\&=Q(-y)
    \\&\le e^{-y^2/2}.
\end{align*}
Thus
\[
0>y\ge -\sqrt{2\ln(4/\eps)}
\]
and so
\[
f_{Z(K)}(y)\ge K p^{1-1/K}\frac{\eps}{4\sqrt{2\pi}}
\ge Kp\frac{\eps}{4\sqrt{2\pi}}\ge \frac{\eps^2}{16\sqrt{2\pi}}.
\]

Now suppose $p^{1/K}\ge 1/2$, so $y\ge 0$. We have
\begin{align*}
    p^{1/K}&=\Phi(y)
    \\&=1-Q(y)
    \\&\ge 1-e^{-y^2/2}
\end{align*}
and so
\[
y\le \sqrt{-2\ln(1-p^{1/K})}.
\]
Thus
\begin{align*}
    f_{Z(K)}(y)&\ge Kp^{1-1/K}\phi\left(\sqrt{-2\ln(1-p^{1/K})}\right)
    \\&=Kp^{1-1/K}\frac{1}{\sqrt{2\pi}} (1-p^{1/K})
    \\&=Kp\frac{1}{\sqrt{2\pi}} (p^{-1/K}-1)
\end{align*}
In the limit as $K\to\infty$,
\begin{align*}
    p^{-1/K}-1
    &=\exp\left\{-\frac{1}{K}\ln p\right\}-1
    \\&\ge-\frac{1}{K}\ln p.
\end{align*}
Thus
\begin{align*}
    f_{Z(K)}(y)&\ge \frac{p\ln(1/p)}{\sqrt{2\pi}}.
\end{align*}
This proves that there exists a $c>0$ such that $f_{Z(K)}(y)\ge c$ for all $y$ in the range of interest. Similar $f_{Z_0}$ is upper and lower bounded as shown above, we may apply Lemma~\ref{lemma:indep_var_pdf} to complete the proof.



\begin{thebibliography}{10}
\providecommand{\url}[1]{#1}
\csname url@samestyle\endcsname
\providecommand{\newblock}{\relax}
\providecommand{\bibinfo}[2]{#2}
\providecommand{\BIBentrySTDinterwordspacing}{\spaceskip=0pt\relax}
\providecommand{\BIBentryALTinterwordstretchfactor}{4}
\providecommand{\BIBentryALTinterwordspacing}{\spaceskip=\fontdimen2\font plus
\BIBentryALTinterwordstretchfactor\fontdimen3\font minus
  \fontdimen4\font\relax}
\providecommand{\BIBforeignlanguage}[2]{{%
\expandafter\ifx\csname l@#1\endcsname\relax
\typeout{** WARNING: IEEEtran.bst: No hyphenation pattern has been}%
\typeout{** loaded for the language `#1'. Using the pattern for}%
\typeout{** the default language instead.}%
\else
\language=\csname l@#1\endcsname
\fi
#2}}
\providecommand{\BIBdecl}{\relax}
\BIBdecl

\bibitem{willems1983discrete}
F.~Willems, ``The discrete memoryless multiple access channel with partially
  cooperating encoders,'' \emph{IEEE Transactions on Information Theory},
  vol.~29, no.~3, pp. 441--445, 1983.

\bibitem{willems1985discrete}
F.~Willems and E.~Van~der Meulen, ``The discrete memoryless multiple-access
  channel with cribbing encoders,'' \emph{IEEE Transactions on Information
  Theory}, vol.~31, no.~3, pp. 313--327, 1985.

\bibitem{noorzad2014power}
P.~Noorzad, M.~Effros, M.~Langberg, and T.~Ho, ``{On the power of cooperation:
  Can a little help a lot?}'' in \emph{IEEE International Symposium on
  Information Theory}, 2014, pp. 3132--3136.

\bibitem{NEL:18}
P.~Noorzad, M.~Effros, and M.~Langberg, ``{The unbounded benefit of encoder
  cooperation for the k-user MAC},'' \emph{IEEE Transactions on Information
  Theory}, vol.~64, no.~5, pp. 3655--3678, 2018.

\bibitem{langberg2016capacity}
M.~Langberg and M.~Effros, ``On the capacity advantage of a single bit,'' in
  \emph{2016 IEEE Globecom Workshops (GC Wkshps)}.\hskip 1em plus 0.5em minus
  0.4em\relax IEEE, 2016, pp. 1--6.

\bibitem{noorzad2018can}
P.~Noorzad, M.~Effros, and M.~Langberg, ``Can negligible cooperation increase
  capacity? the average-error case,'' in \emph{Proceedings of IEEE
  International Symposium on Information Theory (ISIT)}, 2018, pp. 1256--1260.

\bibitem{NLE:19}
P.~Noorzad, M.~Langberg, and M.~Effros, ``{Negligible Cooperation: Contrasting
  the Maximal- and Average-Error Cases},'' \emph{Manuscript. Available on
  https://arxiv.org/pdf/1911.10449.pdf}, 2019.

\bibitem{hartigan2014bounding}
J.~Hartigan \emph{et~al.}, ``Bounding the maximum of dependent random
  variables,'' \emph{Electronic Journal of Statistics}, vol.~8, no.~2, pp.
  3126--3140, 2014.

\bibitem{Borell}
C.~Borell, ``{The Brunn-Minkowski inequality in Gauss space},''
  \emph{Inventiones Mathematicae}, vol.~30, no.~2, pp. 207--216, 1975.

\bibitem{TIS}
B.~Tsirelson, I.~Ibragimov, and V.~Sudakov, ``{Norms of Gaussian sample
  functions},'' \emph{Proceedings of the Third Japan-USSR Symposium on
  Probability Theory}, vol. 550, pp. 20--41, 1976.

\bibitem{huang2012finite}
Y.-W. Huang and P.~Moulin, ``Finite blocklength coding for multiple access
  channels,'' in \emph{2012 IEEE International Symposium on Information Theory
  Proceedings}.\hskip 1em plus 0.5em minus 0.4em\relax IEEE, 2012, pp.
  831--835.

\bibitem{jazi2012simpler}
E.~M. Jazi and J.~N. Laneman, ``Simpler achievable rate regions for multiaccess
  with finite blocklength,'' in \emph{2012 IEEE International Symposium on
  Information Theory Proceedings}.\hskip 1em plus 0.5em minus 0.4em\relax IEEE,
  2012, pp. 36--40.

\bibitem{tan2014dispersions}
V.~Y. Tan and O.~Kosut, ``On the dispersions of three network information
  theory problems,'' \emph{IEEE Transactions on Information Theory}, vol.~60,
  no.~2, pp. 881--903, 2014.

\bibitem{scarlett2015second}
J.~Scarlett, A.~Martinez, and A.~G. i~F{\`a}bregas, ``Second-order rate region
  of constant-composition codes for the multiple-access channel,'' \emph{IEEE
  Transactions on Information Theory}, vol.~61, no.~1, pp. 157--172, 2015.

\bibitem{yavas2020random}
R.~C. {Yavas}, V.~{Kostina}, and M.~{Effros}, ``Random access channel coding in
  the finite blocklength regime,'' \emph{IEEE Transactions on Information
  Theory}, 2020.

\bibitem{chen2013stein}
L.~H.~Y. Chen, X.~Fang, and Q.-M. Shao, ``From {Stein} identities to moderate
  deviations,'' \emph{Ann. Probab.}, vol.~41, no.~1, pp. 262--293, 01 2013.

\end{thebibliography}
\end{document}